\documentclass[journal]{IEEEtran}

\usepackage[cmex10]{amsmath}
\usepackage{array}
\usepackage{url}
\usepackage{amsfonts}
\usepackage{amssymb}

\DeclareMathOperator{\aut}{Aut} \DeclareMathOperator{\trans}{Trans}
\DeclareMathOperator{\supp}{supp} \DeclareMathOperator{\wt}{wt}
\DeclareMathOperator{\diag}{diag} \DeclareMathOperator{\paut}{PAut}
\DeclareMathOperator{\lu}{LU} 
\DeclareMathOperator{\tr}{Tr}

\newtheorem{definition}{Definition}
\newtheorem{example}{Example}
\newtheorem{lemma}{Lemma}
\newtheorem{theorem}{Theorem}
\newtheorem{corollary}{Corollary}
\newtheorem{remark}{Remark}

\begin{document}

\title{Transversality versus Universality \\ for Additive Quantum Codes}

\author{Bei~Zeng, 
        Andrew~Cross, 
        and~Isaac~L.~Chuang
\thanks{The authors are with the Center for Ultracold Atoms, Department
of Physics, Massachusetts Institute of Technology, Cambridge,
MA, 02139 USA e-mail: zengbei@mit.edu.}
}



\maketitle

\begin{abstract}
Certain quantum codes allow logic operations to be performed on the
encoded data, such that a multitude of errors introduced by faulty
gates can be corrected.  An important class of such operations are
{\em transversal}, acting bitwise between corresponding qubits in each
code block, thus allowing error propagation to be carefully limited.
If any quantum operation could be implemented using a set of such
gates, the set would be {\em universal}; codes with such a universal,
transversal gate set have been widely desired for efficient
fault-tolerant quantum computation.  We study the structure of
$GF(4)$-additive quantum codes and prove that no universal set of
transversal logic operations exists for these codes.  This result
strongly supports the idea that additional primitive operations, based
for example on quantum teleportation, are necessary to achieve
universal fault-tolerant computation on additive codes.
\end{abstract}


\section{Introduction}
\IEEEPARstart{T}he
study of fault-tolerant quantum computation is essentially driven
by the properties of quantum codes -- specifically, what logic
operations can be implemented on encoded data, without decoding, and
while controlling error
propagation \cite{shor:ftqc,Preskill98b,gottesman:HR,nielsen:NC}.  Quantum code
automorphisms, and their close relatives, {\em transversal} gates, are
among the most widely used and simplest fault-tolerant logic gates;
uncorrelated faults before and during such gates result in
uncorrelated errors in the multi-qubit blocks.  Transversal gates, in
particular, are gates that act bitwise, such that they may be
represented by tensor product operators in which the $j$th term acts
only on the $j$th qubit from each block \cite{gottesman:thesis}.  Much
like in classical computation, not all gate sets can be composed to
realize an arbitrary operation, however.  It would be very desirable
to find a {\em universal} transversal gate set, from which any quantum
operation could be composed, because this could dramatically simplify
resource requirements for fault-tolerant quantum
computation \cite{Metodi05a,Oskin02}.  In particular, the accuracy
threshold would likely improve, if any quantum computation could be
carried out with transversal gates alone\cite{aliferis:bs}.

Many of the well-known $GF(4)$-additive codes (also known as {\em
stabilizer} codes \cite{crss:gf4,gottesman:thesis}) have been
exhaustively studied, for their suitability for fault tolerant quantum
computation.  However, no quantum code has yet been discovered, which
has automorphisms allowing a universal transversal gate set.
Specifically, an important subset, the {\em CSS} codes
\cite{crss:gf4,steane:ec,steane:fta}, all admit a useful two-qubit
transversal primitive, the controlled-{\sc not} (``CNOT'') gate, but
each {\em CSS} code seems to lack some important element that would fill a universal
set.

For example, the $[[n,k,d]] = [[7,1,3]]$ Steane code \cite{steane:ec},
based on a Hamming code and its dual, has transversal gates generating
the Clifford group.  This group is the finite group of symmetries of
the Pauli group\cite{nielsen:NC}, and may be generated by the CNOT,
the Hadamard, and the single-qubit phase gate.  For the Steane code, a
Clifford gate can be implemented by applying that gate (or its
conjugate) to each coordinate \cite{shor:ftqc}.  Moreover, encoding,
decoding, and error correction circuits for {\em CSS} codes can be
constructed entirely from Clifford operations, and thus Clifford group
gates are highly desirable for efficient fault-tolerant circuits.
Unfortunately, it is well known that gates in the Clifford group are
not universal for quantum computation, as asserted by the
Gottesman-Knill theorem \cite{nielsen:NC,gottesman:HR}.  In fact,
Rains has shown that the automorphism group of any $GF(4)$-linear code
(i.e. the {\em CSS} codes) lies in the Clifford group \cite{rains:d2}.  Because
of this, and also exhaustive searches, it is believed that the Steane
code, which is a $GF(4)$-linear code, does not have a universal set of
transversal gates.

The set of Clifford group gates is not universal, but it is also well
known that the addition of nearly any gate outside of this set (any
``non-Clifford'' gate) can complete a universal
set \cite{nebe:clifford}.  For example, the single-qubit $\pi/8$, or
$T=\diag{(1,e^{i\pi/4})}$ gate, is one of the simplest non-Clifford
gates which has widely been employed in fault-tolerant constructions.
Codes have been sought which allow a transversal $T$ gate.


Since additive codes have a simple structure, closely related to the
abelian subgroup of Pauli groups, transversal Clifford gates for such
codes may be constructed systematically \cite{gottesman:thesis}.
However, how to find non-Clifford transversal gates for a given code
is not generally known.  Some intriguing examples have been
discovered, however.  Strikingly, the $[[15,1,3]]$ {\em CSS} code
constructed from a punctured Reed-Muller code has a transversal $T$
gate \cite{knill:ta}.  Rather frustratingly, however, this code does
not admit a transversal Hadamard gate, thus leaving the Clifford gate
set incomplete, and rendering the set of transversal gates on that
code non-universal.

In fact, all known examples of transversal gate sets on quantum codes
have been deficient in one way or another, leading to
non-universality.  Some of the known $[[n,1,3]]$ code results are
listed in Table~\ref{table1}.  None of these codes listed, or known so
far in the community, allows a universal set of transversal gates.

\begin{table}[htbp]
   \centering
\begin{tabular}{|c|c|c|}
	\hline
\rule{0pt}{2.4ex}
Code & Transversal gates  &   Gates not transversal  \\
	\hline
\rule{0pt}{2.4ex}
      $[[5,1,3]]$          & $PH$, M$_\text{3}$     &  $H$, $P$, CNOT, $T$ \\
\rule{0pt}{2.4ex}
      $[[7,1,3]]$       & $H$, $P$, CNOT  & $T$ \\
\rule{0pt}{2.4ex}
      $[[9,1,3]]$       & CNOT  & $H$, $P$, $T$ \\
\rule{0pt}{2.4ex}
      $[[15,1,3]]$    & $T$, CNOT  &  $H$ \\
\rule{0pt}{2.4ex}
      $[[2^m-1,1,3]]$ &$T_\text{m}$, CNOT & $H$\\
	\hline
\end{tabular}
\caption{\protect\parbox[t]{3.15in}{Collection of some $[[n,1,3]]$ codes and
their properties.  The second column lists allowed transversal gates,
and the third column gives the gates which cannot be transversal on
the corresponding codes. $H$ is the Hadamard gate, $P=\diag(1,i)$ is
the phase gate, $T=\diag{(1,e^{i\pi/4})}$ and
$T_m=\diag{(1,e^{i\pi/2^{m-2}})}$.  For the $[[5,1,3]]$ code, $M_3$ is
a three-qubit Clifford operation (see page 89 of
\protect\cite{gottesman:thesis}). The $[[2^m-1,1,3]]$ code with
transversal $T_m$ gate is a {\em CSS} code constructed from punctured binary
Reed-Muller code $RM^*(1,m)$ and its even
subcode {\protect\cite{zeng:lulc}}.}}
\label{table1}
\end{table}

Considering the many unsuccessful attempts to construct a code with a
universal set of transversal gates, it has been widely conjectured in
the community that transversality and universality on quantum codes
are incompatible; specifically, it is believed that \textit{no
universal set of transversal gates exists, for any quantum code $Q$},
even allowing for the possibility of additional qubit permutation
operations inside code blocks.

Our main result, given in Section~\ref{sec:transversal}, proves a
special case of this ``$T$ versus $U$'' incompatibility, where $Q$ is
a $GF(4)$-additive code and coordinate permutations are not allowed.
Our proof relies on earlier results by Rains \cite{rains:d2} and Van
den Nest \cite{nest:lulc}, generalized to multiple blocks encoded in
additive quantum codes.  In Section~\ref{sec:permutation}, we prove
$T$ vs. $U$ incompatibility for a single block of qubits encoded in a
$GF(4)$-additive code, by clarifying the effect of coordinate
permutations. In Section~\ref{sec:applications}, we consider the
allowable transversal gates on additive codes, using the proof
technique we employ. We also present a simple construction based on
classical divisible codes that yields many quantum codes with
non-Clifford transversal gates on a single block.  We begin in the
next section with some preliminary definitions and terminology.

\section{Preliminaries}

This section reviews definitions and preliminary results about
additive codes \cite{crss:gf4}, Clifford groups and universality,
automorphism groups, and codes stabilized by minimal
elements. Throughout the paper, we only consider $GF(4)$-additive
codes, i.e. codes on qubits, leaving more general codes to future
work. We use the stabilizer language to describe $GF(4)$-additive
quantum codes, which are also called binary stabilizer codes.

\subsection{Stabilizers and stabilizer codes}

\begin{definition}
The \textit{$n$-qubit Pauli group} ${\mathcal G}_n$ consists of all
$4\times 4^n$ operators of the form $R=\alpha_R
R_1\otimes\dots\otimes R_n$, where $\alpha_R\in\{\pm 1,\pm i\}$ is a
phase factor and each $R_i$ is either the $2\times 2$ identity
matrix $I$ or one of the Pauli matrices $X$, $Y$, or $Z$. A
\textit{stabilizer} $\mathcal{S}$ is an abelian subgroup of the
$n$-qubit Pauli group ${\mathcal G}_n$ which does not contain $-I$.
A \textit{support} is a subset of $[n]:=\{1,2,\dots,n\}$. The
support $\supp{(R)}$ of an operator $R\in {\mathcal G}_n$ is the set
of all $i\in [n]$ such that $R_i$ differs from the identity, and the
\textit{weight} $wt(R)$ equals the size $|supp(R)|$ of the support.
The set of elements in $\mathcal{G}_n$ that commute with all
elements of $\mathcal{S}$ is the \textit{centralizer}
$\mathcal{C}(\mathcal{S})$.
\end{definition}

\begin{example}
We have the relation $[XXXX,ZZZZ]=0$ where $XXXX=X^{\otimes 4}$
represents a tensor product of Pauli operators. Consider the
stabilizer $\mathcal{S}=\langle XXXX, ZZZZ\rangle$ where
$\langle\cdot\rangle$ indicates a generating set, so
$$\mathcal{S}=\{ IIII, XXXX, ZZZZ, YYYY\}.$$ \noindent We have
$\supp{(XXXX)}=\{1,2,3,4\}$ and $\wt{(XXXX)}=4$, for example.
Finally, the centralizer is
$\mathcal{C}(\mathcal{S})=\langle\mathcal{S}, ZZII,ZIZI, XIXI,
XXII\rangle$.
\end{example}

A stabilizer consists of $2^m$ Pauli operators for some nonnegative
integer $m\leq n$ and is generated by $m$ independent Pauli
operators. As the operators in a stabilizer are Hermitian and
mutually commuting, they can be diagonalized simultaneously.

\begin{definition}
An $n$-qubit \textit{stabilizer code} $Q$ is the joint eigenspace of
a stabilizer $\mathcal{S}(Q)$,
\begin{equation}
Q=\{ |\psi\rangle\in ({\mathbb C}^2)^{\otimes n}\ |\
R|\psi\rangle=|\psi\rangle,\forall R\in \mathcal{S}(Q)\}
\end{equation}
where each state vector $|\psi\rangle$ is assumed to be normalized.
$Q$ has dimension $2^{n-m}$ and is called an $[[n,k,d]]$ stabilizer
code, where $k=n-m$ is the \textit{number of logical qubits} and $d$
is the \textit{minimum distance}, which is the weight of the minimum
weight element in $\mathcal{C}(\mathcal{S})\setminus \mathcal{S}$.
The code $Q$ can correct errors of weight
$t\leq\lfloor\frac{d-1}{2}\rfloor$.
\end{definition}

\begin{example}
Continuing, we have
\begin{align*}
Q = \textrm{span}\{ & |0000\rangle + |1111\rangle, |0011\rangle+|1100\rangle, \\
& |1010\rangle+|0101\rangle, |1001\rangle+|0110\rangle\}
\end{align*}
so $n=4$, $m=2$, $\dim{Q}=4$, and $k=2$. From
$C(\mathcal{S})\setminus \mathcal{S}$, we see that $d=2$. Therefore,
$Q$ is a $[[4,2,2]]$ code.
\end{example}

Each set of $n$ mutually commuting independent elements of
$\mathcal{C}(\mathcal{S})$ stabilizes a quantum codeword and
generates an abelian subgroup of the centralizer. This leads to the
isomorphism $\mathcal{C}(\mathcal{S})/\mathcal{S}\cong
\mathcal{G}_k$ that maps each element $X_i,Z_i\in\mathcal{G}_k$ to
a coset representative $\bar{X}_i,\bar{Z}_i\in
C(\mathcal{S})/\mathcal{S}$ \cite{gottesman:thesis}. The isomorphism
associates the $k$ logical qubits to logical Pauli operators
$\bar{X}_i,\bar{Z}_i$ for $i=1,\ldots,k$, and these operators obey
the commutation relations of $\mathcal{G}_k$.

\begin{example}
One choice of logical Pauli operators for the $[[4,2,2]]$ code is
$\bar{X}_1=XIXI$, $\bar{Z}_1=ZZII$, $\bar{X}_2=XXII$, and
$\bar{Z}_2=ZIZI$. These satisfy the commutation relations of
${\mathcal G}_2$.
\end{example}

\subsection{Universality}

Stabilizer codes are stabilized by subgroups of the Pauli group, so
some unitary operations that map the Pauli group to itself also map
the stabilizer to itself, preserving the code space.

\begin{definition}
The \textit{$n$-qubit Clifford group} ${\mathcal L}_n$ is the group
of unitary operations that map ${\mathcal G}_n$ to itself under
conjugation. One way to specify a gate in ${\mathcal L}_n$ is to
give the image of a generating set of ${\mathcal G}_n$ under that
gate. ${\mathcal L}_n$ is generated by the single qubit Hadamard
gate,
\begin{equation}
H: (X,Z)\rightarrow (Z,X),
\end{equation}
the single qubit Phase gate
\begin{equation}
P: (X,Z)\rightarrow (-Y,Z),
\end{equation}
and the two-qubit controlled-not gate
\begin{align}
\textrm{CNOT}: & (XI,IX,XX,ZI,IZ,ZZ)\rightarrow \\ &
(XX,IX,XI,ZI,ZZ,IZ)
\end{align}
by the Gottesman-Knill theorem \cite{nielsen:NC,gottesman:HR}.
\end{definition}

\begin{definition}\label{def:universality}
A set of unitary gates $G$ is (quantum) \textit{computationally
universal} if for any $n$, any unitary operation $U\in SU(2^n)$ can
be approximated to arbitrary accuracy $\epsilon$ in the sup operator
norm $||\cdot||$ by a product of gates in $G$. In notation, $\forall
\epsilon>0, \exists V=V_1V_2\dots V_{\eta(\epsilon)}\ \textrm{where
each}\ V_i\in G\ \textrm{s.t.}\ ||V-U||<\epsilon.$ In this
definition, gates in $G$ may be implicitly mapped to isometries on
the appropriate $2^n$-dimensional Hilbert space.
\end{definition}

The Gottesman-Knill theorem asserts that any set of Clifford group
gates can be classically simulated and is therefore not (quantum)
computationally universal. Quantum teleportation is one technique for
circumventing this limit and
constructing computationally universal sets of gates using Clifford
group gates and measurements of Pauli operators
\cite{gottesman:teleport,zhou:ft}. There is a large set of gates that
arise in fault-tolerant quantum computing through quantum
teleportation.

\begin{definition}
The \textit{${\mathcal C}_k^{(n)}$ hierarchy} is a set of gates that
can be achieved through quantum teleportation and is defined
recursively as follows: ${\mathcal C}_1^{(n)}={\mathcal G}_n$ and
\begin{equation}
{\mathcal C}_k^{(n)} = \{ U\in SU(2^n)\ |\ UgU^\dag\in {\mathcal
C}_{k-1}^{(n)}\ \forall\ g\in {\mathcal C}_1^{(n)}\},
\end{equation}
for $k>1$. ${\mathcal C}_k^{(n)}$ is a group only for $k=1$ and
$k=2$ and ${\mathcal C}_2^{(n)}={\mathcal L}_n$.\label{def:ck}
\end{definition}

The Clifford group generators $\{H,P,CNOT\}$ plus any other gate
outside of the Clifford group is computationally universal
\cite{nebe:clifford}. For example, the gates
$T=\diag{(1,e^{i\pi/4})}\in {\mathcal C}_3^{(1)}\setminus {\mathcal
C}_2^{(1)}$ and $\textsc{Toffoli}\in {\mathcal C}_3^{(3)}\setminus
{\mathcal C}_2^{(3)}$ are computationally universal when taken
together with the Clifford group.

\subsection{Automorphisms of stabilizer codes}

An automorphism is a one-to-one, onto map from some domain back to
itself that preserves a particular structure of the domain. We are
interested in quantum code automorphisms, unitary maps that preserve
the code subspace and respect a fixed tensor product decomposition
of the $n$-qubit Hilbert space. The weight distribution of an
arbitrary operator with respect to the Pauli error basis ${\mathcal
G}_n$ is invariant under these maps. With respect to the tensor
product decomposition, we can assign each qubit a coordinate $j\in
[n]$, in which case the quantum code automorphisms are those local
operations and coordinate permutations that correspond to logical
gates. In some cases, these automorphisms correspond to the
permutation, monomial, and/or field automorphisms of classical codes
\cite{huffman:HP}. This section formally defines logical gates and
quantum code automorphisms on an encoded block.

\begin{definition}
A unitary gate $U\in SU(2^n)$ acting on $n$ qubits is a
\textit{logical gate on $Q$} if $[U,P_Q]=0$
where $P_Q$ is the orthogonal projector onto $Q$ given by
\begin{equation}
P_Q = \frac{1}{2^m}\sum_{R\in \mathcal{S}(Q)}R.
\end{equation}
Let ${\mathbb V}(Q)$ denote the set of logical gates on $Q$. When
$Q$ is understood, we simply say that the gate $U$ is a logical
gate. The logical gates ${\mathbb V}(Q)$ are a group that is
homomorphic to $SU(2^k)$ since it is possible to encode an arbitrary
$k$-qubit state in the code.
\end{definition}

\begin{example}
For the $[[4,2,2]]$ code, $P_Q=\frac{1}{4}(I^{\otimes 4}+X^{\otimes
4}+Y^{\otimes 4}+Z^{\otimes 4})$. Any unitary acting in the code
manifold
\begin{align*}
\alpha (|0000\rangle+|1111\rangle) + \beta(|0011\rangle+|1100\rangle) + \\
\gamma(|1010\rangle+|0101\rangle) +
\delta(|1001\rangle+|0110\rangle)
\end{align*}
is a logical gate.
\end{example}

\begin{definition}
The \textit{full automorphism group} $\aut{(Q)}$ of $Q$ is the collection of all
logical operations on $Q$ of the form $P_\pi U$ where $P_\pi$ enacts
the coordinate permutation $\pi$ and $U=U_1\otimes\dots\otimes U_n$
is a local unitary operation. The product of two such operations is
a logical operation of the same form, and operations of this form
are clearly invertible, so $\aut{(Q)}$ is indeed a group. More
formally, the full automorphism group $\aut{(Q)}$ of $Q$ is sometimes
defined as the subgroup of logical operations contained in the semidirect product
$(S_n,SU(2)^{\otimes n},\nu)$, where
$\nu:S_n\rightarrow\aut{(SU(2)^{\otimes n})}$ is given by
\begin{equation}
\nu(\pi)(U_1\otimes\dots\otimes U_n)=U_{\pi(1)}\otimes\dots\otimes
U_{\pi(n)}
\end{equation}
and $S_n$ is the symmetric group of permutations on $n$ items. The
notation $S_n\ltimes SU(2)^{\otimes n}$ is sometimes used. When
$\aut{(Q)}$ is considered as a semidirect product group, an element
$(\pi,U_1\otimes\dots\otimes U_n)\in\aut{(Q)}$ acts on codewords as
$U_1\otimes\dots\otimes U_n$ and on coordinate labels as $\pi$. The
product of two automorphisms in $\aut{(Q)}$ is
\begin{equation}
(\pi_1,U)(\pi_2,V)=(\pi_1\pi_2,(U_{\pi_2(1)}V_1)\otimes\dots\otimes
(U_{\pi_2(n)}V_n)),
\end{equation}
by definition of the semidirect product.
\end{definition}

The full automorphism group contains several interesting subgroups.
Consider the logical gates that are local
\begin{equation}
\lu{(Q)} = \{ U\in {\mathbb V}(Q)\ |\ U=\otimes_{i=1}^n U_i,\ U_i\in
SU(2)\}
\end{equation}
and the logical gates that are implemented by permutations
\begin{equation}
\paut{(Q)} = \{ \pi\in S_n\ |\ P_\pi\in {\mathbb V}(Q)\}
\end{equation}
where $P_\pi:S_n\rightarrow SU(2^n)$ is defined by
$P_\pi|\psi_1\psi_2\dots\psi_n\rangle=|\psi_{\pi(1)}\psi_{\pi(2)}\dots\psi_{\pi(n)}\rangle$
on the computational basis states. The semidirect product of these
groups is contained in the full automorphism group, i.e.
$\paut{(Q)}\ltimes\lu{(Q)}\subseteq \aut{(Q)}$. In other words, the
elements of this subgroup are products of automorphisms for which
either $P_\pi=I$ or $U=I$, in the notation of the definition. In
general, $\aut{(Q)}$ may be strictly larger than
$\paut{(Q)}\ltimes\lu{(Q)}$, as happens with the family of
Bacon-Shor codes \cite{aliferis:bs}. The automorphism group of $Q$
as a $GF(4)$-additive classical code is a subgroup of the full
automorphism group, since classical automorphisms give rise to
quantum automorphisms in the Clifford group.

\begin{example}
For the $[[4,2,2]]$, $\lu{(Q)}=\langle P^{\otimes 4},H^{\otimes
4}\rangle\cong S_3$ and $\paut{(Q)}=S_4$. Furthermore, the
full automorphism group $\aut{(Q)}=S_4\times S_3$ equals the
automorphism group of the $[[4,2,2]]$ as a $GF(4)$-additive code and
$\paut{(Q)}\ltimes\lu{(Q)}=\aut{(Q)}$ \cite{rains:d2}.
\end{example}

\subsection{Fault-tolerance and multiple encoded blocks}

As we alluded to earlier, the reason we find $\aut{(Q)}$ interesting is 
because gates in $\aut{(Q)}$ are ``automatically'' fault-tolerant. Fault-tolerant
gate failure rates are at least quadratically suppressed after error-correction. 
Given some positive
integer $t'\leq t$, two properties are sufficient (but not
necessary) for a gate to be fault-tolerant. First, the gate must
take a weight $w$ Pauli operator, $0\leq w\leq t'$, to a Pauli
operator with weight no greater than $w$ under conjugation. Second,
if $w$ unitaries in the tensor product decomposition of the gate are
replaced by arbitrary quantum operations acting on the same qubits,
then the output deviates from the ideal output by the action of an
operator with weight no more than $w$. Gates in $\aut{(Q)}$ have
these properties for any $t'\in [n]$ if we consider the permutations
to be applied to the qubit labels rather than the quantum state.

We are also interested in applying logic gates between multiple
encoded blocks so that it is possible to simulate a large logical computation using
any stabilizer code we choose. In general, each block can be encoded in a different
code. Logic gates between these blocks could take inputs encoded in
one code to outputs encoded in another, as happens with some logical
gates on the polynomial codes \cite{abo:ft} or with code
teleportation \cite{gottesman:teleport}.

In this paper, we only consider the simplest situation where blocks
are encoded using the same code and gates do not map between codes.
Our multiblock case with $r$ blocks has $rk$ qubits encoded in the
code $Q^{\otimes r}$ for some positive integer $r$. The notion of a
logical gate is unchanged for the multiblock case: $Q$ is replaced
by $Q^{\otimes r}$ in the prior definitions. However, the
fault-tolerance requirements become: (1) a Pauli operator with weight
$w_i$ on input block $i$ and $\sum_i w_i\leq t'$ conjugates to a
Pauli operator with weight no greater than $\sum_i w_i$ on
\textit{each} output block and (2) if $w\leq t'$ unitaries in the
tensor product decomposition of the gate are replaced by arbitrary
quantum operations acting on the same qubits, then \textit{each}
output block may deviate from the ideal output by no more than a weight
$w$ operator.

Gates in $\aut{(Q^{\otimes r})}$ are also fault-tolerant, since the
only new behavior comes from the fact that $\paut{(Q)}^{\otimes r}$
is not generally equal to $\paut{(Q^{\otimes r})}$. However,
$\aut{(Q^{\otimes r})}$ does not contain all of the fault-tolerant
gates on $r$ blocks because we can interact qubits in different
blocks and still satisfy the fault-tolerance properties.

\begin{definition}\label{def:transversal}
A \textit{transversal $r$-qubit gate} on $Q^{\otimes r}$ is a
unitary gate $U\in{\mathbb V}(Q^{\otimes r})$ such that
\begin{equation}\label{eq:trans}
U = \otimes_{j=1}^n U_j,
\end{equation}
where $U_j\in SU(2^r)$ only acts on the $j$th qubit of each block.
Let $\trans{(Q^{\otimes r})}$ denote the $r$-qubit transversal
gates.
\end{definition}

More generally, we could extend the definition of transversality to
allow coordinate permutations, as in the case of code automorphisms,
and still satisfy the fault-tolerance properties given above.
However, we keep the usual definition of transversality and do not 
make this extension here.

\subsection{Codes stabilized by minimal elements and the minimal support condition}\label{subsec:minsupp}

\begin{definition}
A \textit{minimal support} of $\mathcal{S}(Q)$ is a nonempty set
$\omega\subseteq [n]$ such that there exists an element in
$\mathcal{S}(Q)$ with support $\omega$, but no elements exist with
support strictly contained in $\omega$ (excluding the identity
element, whose support is the empty set). An element in $\mathcal{S}(Q)$ with minimal support is
called a \textit{minimal element}. For each minimal support
$\omega$, let $\mathcal{S}_{\omega}(Q)$ denote the stabilizer
generated by minimal elements with support $\omega$ and let
$Q_\omega$ denote the \textit{minimal code associated to $\omega$},
stabilized by $\mathcal{S}_{\omega}(Q)$. Let $\mathcal{M}(Q)$ denote
the \textit{minimal support subgroup} generated by all minimal
elements in $\mathcal{S}(Q)$.
\end{definition}

\begin{example}
Consider the $[[5,1,3]]$ code $Q$ whose stabilizer is generated by
$XZZXI$ and its cyclic shifts. Every set of 4 contiguous coordinates
modulo the boundary is a minimal support: $\{1,2,3,4\}$,
$\{2,3,4,5\}$, $\{3,4,5,1\}$, etc. The minimal elements with support
$\omega=\{1,2,3,4\}$ are $XZZXI$, $YXXYI$, and $ZYYZI$. Therefore,
the minimal code $Q_\omega$ is stabilized by
$\mathcal{S}_{\omega}(Q) = \langle XZZXI, YXXYI \rangle$. This code
is a $[[4,2,2]]\otimes [[1,1,1]]$ code, since this $[[4,2,2]]$ code
is locally equivalent to the code stabilized by $\langle XXXX,
ZZZZ\rangle$ by the equivalence $I\otimes C\otimes C\otimes I$,
where $C:X\mapsto Y\mapsto Z\mapsto X$ by conjugation. The
$[[5,1,3]]$ code is the intersection of its minimal codes, meaning
$Q=\cap_\omega Q_\omega$ and $\mathcal{S}(Q)=\prod_\omega
\mathcal{S}_\omega(Q)$ where the intersection and product run over
the minimal supports. Furthermore, $\mathcal{M}(Q)=\mathcal{S}(Q)$.
\end{example}

Given an arbitrary support $\omega$, the projector
$\rho_{\omega}(Q)$ obtained by taking the partial trace of $P_Q$
over $\bar{\omega}:=[n]\setminus\omega$ is
\begin{equation}\label{omega}
\rho_{\omega}(Q) = \frac{1}{B_{\omega}(Q)}\sum_{R\in \mathcal{S}(Q),
supp(R)\subseteq\omega} \tr_{\bar{\omega}} R,
\end{equation}
where $B_\omega(Q)$ is the number of elements of $S$ with support
contained in $\omega$ including the identity. The projector
$\rho_{\omega}(Q)\otimes I_{\bar{\omega}}$ projects onto a subcode $Q_\omega$ of $Q$, $Q\subseteq Q_\omega$, that is
stabilized by the subgroup $\mathcal{S}_{\omega}(Q)$ of
$\mathcal{S}(Q)$.

\begin{example}
For the $[[5,1,3]]$,
$\rho_{\{1,2,3,4\}}=\frac{1}{4}(IIII+XZZX+YXXY+ZYYZ)\cong
P_{[[4,2,2]]}$.
\end{example}

\begin{definition}
If $Q$, $Q'$ are stabilizer codes, a gate $U=U_1\otimes\dots\otimes
U_n$ satisfying $U|\psi\rangle=|\psi'\rangle\in Q'$ for all
$|\psi\rangle\in Q$ is a \textit{local unitary (LU) equivalence from
$Q$ to $Q'$} and $Q$ and $Q'$ are called \textit{locally equivalent
codes}. If each $U_i\in {\mathcal L}_1$ then $Q$ and $Q'$ are
called \textit{locally Clifford equivalent codes} and $U$ is a
\textit{local Clifford (LC) equivalence from $Q$ to $Q'$}. In this
paper, we sometimes use these terms when referring to the projectors
onto the codes as well.
\end{definition}

The following results are applied in Section~\ref{sec:transversal}.

\begin{lemma}[\cite{rains:d2}]\label{lem:rains}
Let $Q$ be a stabilizer code. If $U=U_1\otimes\dots\otimes U_n$ is a
logical gate for $Q$ then $[U_\omega,\rho_\omega(Q)]=0$ for all
$\omega$, where $U_\omega=\otimes_{i\in\omega} U_i$. More generally,
if $Q'$ is another stabilizer code and $U$ is a local equivalence
from $Q$ to $Q'$ then
\begin{equation}
U_\omega\rho_\omega(Q)U_\omega^\dag=\rho_\omega(Q')\label{Uomega}
\end{equation}
for all $\omega$.
\end{lemma}

\begin{IEEEproof}
$U$ is a local gate, so
\begin{equation}
\tr_{\bar{\omega}} UP_QU^\dag = U_\omega(\tr_{\bar{\omega}}
P_Q)U_\omega^\dag = U_\omega\rho_\omega(Q)U_\omega^\dag.
\end{equation}
Since $U$ maps from $Q$ to $Q'$, we obtain the result.
\end{IEEEproof}

By examining subcodes, we can determine if a given gate can be a
logical gate using Lemma~\ref{lem:rains}.
In particular, if $U$ is not a logical gate for each minimal code of
$Q$, then $U$ cannot be a logical gate for $Q$. 

\begin{definition}
A stabilizer code is called \textit{free of Bell pairs} if it cannot
be written as a tensor product of a stabilizer code and a
$[[2,0,2]]$ code (a Bell pair). A stabilizer code $\cal{S}$ is called
\textit{free of trivially encoded qubits} if for each $j\in[n]$ there 
exists an element $s\in{\cal S}$ such that the $j$th coordinate of
$s$ is not the identity matrix, i.e. if $S$ cannot be written as a
tensor product of a stabilizer code and a $[[1,1,1]]$ code (a trivially
encoded qubit).
\end{definition}

Let ${\mathfrak m}(Q)$ be the union of the minimal supports of a
stabilizer code $Q$. The following theorem is a major tool in the
solution of our main problem.

\begin{theorem}[\cite{rains:d2,nest:lulc}]\label{thm:d2lulc}
Let $Q$, $Q'$ be $[[n,k,d]]$ stabilizer codes, not necessarily distinct,
that are free of Bell pairs and trivially encoded qubits, and let
$j\in {\mathfrak m}(Q)$. Then any local equivalence $U$ from $Q$
to $Q'$ must have either $U_j\in {\mathcal L}_1$ or
$U_j=Le^{i\theta R}$ for some $L\in {\mathcal L}_1$, some angle
$\theta$, and some $R\in {\cal G}_1$.
\end{theorem}

\begin{IEEEproof}
For completeness, we include a proof of this theorem here, though it
can be found expressed using slightly different language in \cite{rains:d2}, \cite{nest:lulc}, 
and \cite{nest:thesis}. The proof requires several results about the minimal subcodes 
of a stabilizer code that we present as Lemmas within the proof body. The 
first of these results shows that each minimal subcode is either a 
quantum error-detecting code or a ``classical'' code with a single parity check, neglecting
the $[[|\bar{\omega}|,|\bar{\omega}|,1]]$ part of the space.

\begin{lemma}\label{lem:Aw}
Let $A_\omega(Q)$ denote the cardinality of the set of elements $s\in {\cal S}$ with support
$\omega$ and let $Q$ be a stabilizer code with stabilizer $\cal S$. If
$\omega$ is a minimal support of $\cal S$, then exactly one of the
following is true:
\begin{enumerate}
\item[(i)] $A_\omega(Q)=1$ and $\rho_\omega(Q)$ is locally Clifford equivalent to
\begin{equation}
\rho_{[[|\omega|,|\omega|-1,1]]}:=\frac{1}{2^{|\omega|}}(I^{\otimes|\omega|}+Z^{\otimes|\omega|}),
\end{equation}
a projector onto a $[[|\omega|,|\omega|-1,1]]$ stabilizer code
$Q_{[[|\omega|,|\omega|-1,1]]}$.
\item[(ii)] $A_\omega(Q)=3$, $|\omega|$ is even, and $\rho_\omega(Q)$ is locally Clifford equivalent to
\begin{align*}
\rho_{[[2m,2m-2,2]]}:=\frac{1}{2^{|\omega|}} & (I^{\otimes|\omega|}+X^{\otimes|\omega|} \\
 & +(-1)^{|\omega|/2}Y^{\otimes|\omega|}+Z^{\otimes|\omega|}),
\end{align*}
a projector onto a $[[2m,2m-2,2]]$ stabilizer code
$Q_{[[2m,2m-2,2]]}$, $m=|\omega|/2$.
\end{enumerate}
\end{lemma}

\begin{IEEEproof}
For any minimal support $\omega$, $A_\omega(Q)\geq 1$. If
$A_\omega(Q)=1$ then $\mathcal{S}_\omega(Q)$ is generated by a
single element and we are done. If $A_\omega(Q)\geq 2$, let
$M_1,M_2\in \mathcal{S}_\omega(Q)\setminus\{I\}$ be distinct
elements. These elements must satisfy $I\neq (M_1)_j\neq (M_2)_j\neq
I$ for all $j\in\omega$, otherwise $\supp(M_1M_2)$ is strictly
contained in $\omega$, contradicting the fact that $\omega$ is a
minimal support. It follows that $\supp(M_1M_2)=\omega$ and
$\{(M_1)_j,(M_2)_j,(M_1M_2)_j\}$ equals $\{X,Y,Z\}$ up to phase for
all $j\in\omega$. Therefore, $I$, $M_1$, $M_2$, and $M_1M_2$ are the
only elements in ${\cal S}_\omega(Q)$. Indeed, suppose there exists a
fourth element $N\in {\cal S}_\omega(Q)$. Fixing any $j_0\in\omega$, either
$(M_1)_{j_0}$, $(M_2)_{j_0}$, or $(M_1M_2)_{j_0}$ equals $N_{j_0}$,
say $(M_1)_{j_0}=N_{j_0}$. Then, $\supp(M_1N)$ is strictly contained
in $\omega$, a contradiction. Therefore, if $A_\omega(Q)\geq 2$ then
$A_\omega(Q)=3$. The number of coordinates in the support $|\omega|$
must be even since $M_1$ and $M_2$ commute.
\end{IEEEproof}

The next result shows that any local equivalence between two
$[[2m,2m-2,2]]$ stabilizer codes with the same $m\geq 2$ must be a
local Clifford equivalence. In the $m=1$ special case, we have a
$[[2,0,2]]$ code, i.e. a Bell pair locally Clifford equivalent to
the state $(|00\rangle+|11\rangle)/\sqrt{2}$, for which the result
does not hold because $V\otimes V^\ast$ is a local equivalence of
the $[[2,0,2]]$ for any $V\in SU(2)$. This special case is the
reason for introducing the definition of a stabilizer code that is
free of Bell pairs.

\begin{lemma}\label{lem:d2LC}
Fix $m\geq 2$ and let $Q$, $Q'$ be stabilizer codes that are LC
equivalent to $Q_{[[2m,2m-2,2]]}$. If $U\in U(2)^{\otimes 2m}$ is a
local equivalence from $Q$ to $Q'$ then $U\in {\mathcal L}_{2m}$.
\end{lemma}

\begin{IEEEproof}
We must show that every $U\in U(2)^{\otimes 2m}$ satisfying
$U\rho_{[[2m,2m-2,2]]}U^\dag=\rho_{[[2m,2m-2,2]]}$ is a local
Clifford operator. Recall that any 1-qubit unitary operator $V\in
U(2)$ acts on the Pauli matrices as
\begin{equation*}
\sigma_a \mapsto V\sigma_aV^\dag = o_{ax}X + o_{ay}Y + o_{az}Z,
\end{equation*}
for every $a\in \{x,y,z\}$ and where $(o_{ab})\in SO(3)$. In the
standard basis $\{|0\rangle,|1\rangle,|2\rangle\}$ of
${\mathbb R}^3$, the matrix
\begin{equation}
X^{\otimes 2m} + (-1)^m Y^{\otimes 2m}+Z^{\otimes 2m}
\end{equation}
is associated to the vector
\begin{equation}
v := |00\dots 0\rangle + (-1)^m |11\dots 1\rangle + |22\dots
2\rangle\in ({\mathbb R}^3)^{\otimes 2m}
\end{equation}
acted on by $SO(3)^{\otimes 2m}$. We must show that every
$O=O_1\otimes\dots\otimes O_{2m}\in SO(3)^{\otimes 2m}$ satisfying
$Ov=v$ is such that each $O_i$ is a monomial matrix (see
\cite{huffman:HP}; a matrix is monomial if it is the product of a
permutation matrix and a diagonal matrix).

Consider the single qutrit operator
\begin{equation}\label{eq:rainstr}
\langle 0|_1 \tr_{\{3,4,\dots,2m\}}(vv^T)|0\rangle_1,
\end{equation}
acting on the second qutrit (second copy of ${\mathbb R}^3$). The
matrix $vv^T$ has 9 nonzero elements, and the partial trace over the
last $2m-2$ qutrits gives
\begin{equation}
\tr_{\{3,4,\dots,2m\}}(vv^T) = |00\rangle\langle 00| +
|11\rangle\langle 11| + |22\rangle\langle 22|.
\end{equation}
Hence the matrix in Eq.~\ref{eq:rainstr} equals the rank one
projector $|0\rangle\langle 0|$. Therefore, if $Ov=v$ then the
operator
\begin{equation}
\langle 0|_1 \tr_{\{3,4,\dots,2m\}} (Ovv^TO^T)|0\rangle_1
\end{equation}
equals $|0\rangle\langle 0|$ as well. The operator is given by the
matrix
\begin{align}
O_2\langle 0|_1(O_1\otimes I)\tr_{\{3,4,\dots,2m\}}(vv^T)(O_1^T\otimes I)|0\rangle_1 O_2^T \\
= O_2\left(\begin{array}{ccc} (O_1)_{00}^2 & 0 & 0 \\ 0 &
(O_1)_{01}^2 & 0 \\ 0 & 0 & (O_1)_{02}^2 \end{array}\right)O_2^T.
\label{eq:monmatrix}
\end{align}
where we have factored $O_2$ to the outside. The matrix within
Eq.~\ref{eq:monmatrix} equals the rank one projector
$O_2^T|0\rangle\langle 0|O_2$ if and only if exactly one of the
elements $(O_1)_{00}$, $(O_1)_{01}$, or $(O_1)_{02}$ is nonzero.
Repeating the argument for every row of $O_1$ by considering the
operators $\langle i|_1\tr_{\{3,4,\dots,2m\}}(Ovv^TO^T)|i\rangle_1$,
$i\in\{0,1,2\}$, shows that every row of $O_1$ has exactly one
nonzero entry. $O_1$ is nonsingular therefore $O_1$ is a monomial
matrix. The vector $v$ is symmetric so repeating the analogous
argument for each operator $O_i$, $i\in [2m]$, completes the proof.
\end{IEEEproof}

Now we can complete the proof of Theorem~\ref{thm:d2lulc}. Let $Q$,
$Q'$ be stabilizer codes, let $U$ be a local equivalence from $Q$ to
$Q'$, and take a coordinate $j\in {\mathfrak m}(Q)$. There is a least 
one element $M\in \mathcal{M}(Q)$ with $j\in\omega:=\supp(M)$. 
Either $A_\omega(Q)=1$ or $A_\omega(Q)=3$ by Lemma~\ref{lem:Aw}.

If $A_\omega(Q)=3$ then $\rho_\omega(Q)$ is LC equivalent to
$\rho_{[[|\omega|,|\omega|-2,2]]}$.
Moreover, as $Q$ is locally equivalent to $Q'$,
$\omega$ is also a minimal support of $\mathcal{S}(Q')$ with
$A_\omega(Q')=3$. Therefore, $\rho_\omega(Q')$ is local Clifford equivalent to
$\rho_{[[|\omega|,|\omega|-2,2]]}$. By Lemma~\ref{lem:rains},
$U_\omega$ maps $\rho_\omega(Q)$ to $\rho_\omega(Q')$ under
conjugation. Note that we must have $|\omega|>2$, otherwise $Q$ is
not free of Bell pairs. Since $|\omega|$ is even, $|\omega|\geq 4$.
By Lemma~\ref{lem:d2LC}, $U_\omega\in {\mathcal L}_{|\omega|}$ so
$U_j\in {\mathcal L}_1$.

If $A_\omega(Q)=1$ and there are elements $R_1,R_2,R_3\in
\mathcal{M}(Q)$ such that $(R_1)_j=X$, $(R_2)_j=Y$, and $(R_3)_j=Z$,
then there exists another minimal element $N\in \mathcal{M}(Q)$ such
that $j\in\mu:=\supp(N)$ and $M_j \neq N_j$. If $A_\mu(Q)=3$ then we
can apply the previous argument to conclude that $U_j\in {\mathcal
L}_1$. Otherwise, $A_\mu(Q)=1$ and
\begin{align}\label{eq:Aw11}
\rho_\omega(Q) & = \frac{1}{2^{|\omega|}}(I^{\otimes|\omega|}+M_\omega) \\
\rho_\mu(Q) & = \frac{1}{2^{|\mu|}}(I^{\otimes|\mu|}+N_\mu).
\end{align}
Since $\omega$ and $\mu$ are also minimal supports of
$\mathcal{S}(Q')$ with $A_\omega(Q')=1$ and $A_\mu(Q')=1$, there
exist unique $M',N'\in S(Q')$ such that
\begin{align}
\rho_\omega(Q') & = \frac{1}{2^{|\omega|}}(I^{\otimes|\omega|}+M'_\omega) \\
\rho_\mu(Q') & =
\frac{1}{2^{|\mu|}}(I^{\otimes|\mu|}+N'_\omega).\label{eq:Aw12}
\end{align}
Applying Lemma~\ref{lem:rains} to $U_\mu$ and $U_\nu$, we have
\begin{align}
U_jM_jU_j^\dag & = \pm M_j' \\
U_jN_jU_j^\dag & = \pm N_j'
\end{align}
from Eqs.~\ref{eq:Aw11}-\ref{eq:Aw12}. These identities show that
$U_j\in {\mathcal L}_1$.

Finally, if $A_\omega(Q)=1$ and $R=(R_1)_j=(R_2)_j$ for any
$R_1,R_2\in \mathcal{M}(Q)$ then {\em any} minimal support $\mu$
such that $j\in\mu$ satisfies $A_\mu(Q)=1$. Applying
Lemma~\ref{lem:rains} to $U_\mu$, we have
\begin{equation}
U_jRU_j^\dag = \pm R'
\end{equation}
for some $R'\in\{X,Y,Z\}$. Choose $L\in {\mathcal L}_1$ such that
$LRL^\dag=R'$. Then $U_j=Le^{i\theta R}$ and the proof of
Theorem~\ref{thm:d2lulc} is complete.
\end{IEEEproof}

\subsection{Coordinates not covered by minimal supports}

It is not always the case the ${\mathfrak m}(Q)=[n]$, as the following example shows.

\begin{example}\label{em:nonminimal}
Consider a $[[6,2,2]]$ code with stabilizer $\mathcal{S}=\langle XXXXII,ZZIIZZ,IIIIXX,IIXXZZ\rangle$. For $j=1,2$, there is no 
minimal support $\omega$ of $\mathcal{S}$ such that $j\in\omega$ \cite{bravyi:thanks}.
\end{example}

For coordinates which are not covered by minimal supports of $\mathcal{S}$, the results in Sec.~\ref{subsec:minsupp} tell us nothing 
about the allowable form of $U_j$ for a transversal gate $U=\bigotimes_{j=1}^n U_j$, so we need another approach for these coordinates.

Let $\mathcal{S}_j=\{ R\ |\ R\in \mathcal{S}(Q),\ j\in\supp(R)\}$ and define the ``minimal elements'' of this set to be
$\mathcal{M}(\mathcal{S}_j)=\{ R\in\mathcal{S}_j\ |\ \nexists R'\in\mathcal{S}_j\ \textrm{s.t.}\ \supp(R')\subsetneq\supp(R)\}$. Note that these sets
do not define codes because they are not necessarily groups.

\begin{lemma}\label{lem:neqsupp}
If $j$ is not contained in any minimal support of $\mathcal{S}$, then for any $R,R'\in\mathcal{M}(\mathcal{S}_j)$ such that the $j$th coordinates
satisfy $R|_j\neq R'|_j$, we must have $\supp(R)\neq\supp(R')$.
\end{lemma}

\begin{IEEEproof} 
We prove by contradiction. If there exist $R,R'\in\mathcal{M}(\mathcal{S}_j)$ such that $R|_j\neq R'|_j$ and $\supp(R)=\supp(R')=\omega$, then 
up to a local Clifford operation, we have $R=X^{\otimes|\omega|}$ and $R'=Z^{\otimes|\omega|}$. Without loss of generality, assume $j=1$. 
Since $\omega$ is minimal in $\mathcal{S}_j$ but not minimal in $\mathcal{S}$, there exists an element $F$ in $\mathcal{S}\setminus\mathcal{S}_j$ 
whose support $\supp(F)=\omega'$ is strictly contained in $\omega$, i.e. $\omega'\subsetneq\omega$. Since $F$ is not in $\mathcal{S}_j$,
$RF$, $R'F$, $R'RF\in\mathcal{M}(\mathcal{S}_j)$. However, one of $RF$, $R'F$, $R'RF\in\mathcal{M}(\mathcal{S}_j)$ will have support that 
is strictly contained in $\omega$, contradicting the fact that $\omega$ is a minimal support of $\mathcal{S}_j$.
\end{IEEEproof}

\begin{lemma}\label{lem:ujnsupp}
If $j$ is not contained in any minimal support of $\mathcal{S}$, then for any transversal gate $U=\bigotimes_{j=1}^n U_j$, one of the following three 
relations is true: $U_jX_jU_j=\pm X_j$, $U_jY_jU_j=\pm Y_j$, $U_jZ_jU_j=\pm Z_j$. In other words, $U_j=Le^{i\theta R}$ for some $L\in {\mathcal L}_1$, 
some angle $\theta$, and some $R\in {\cal G}_1$.
\end{lemma}

\begin{IEEEproof}
For any element $R\in\mathcal{M}({S}_j)$ with a fixed support $\omega$, we have $R|_j=Z$ up to local Clifford operations by Lemma~\ref{lem:neqsupp}. 
Tracing out all the qubits in $\bar{\omega}$, we get a reduced density matrix $\rho_{\omega}$ with the form
\begin{equation}\label{eq:rhonsupp}
\rho_{\omega}=\frac{1}{2^{|\omega|}}(I_j\otimes R_I+Z_j\otimes R_Z),
\end{equation}
where $R_I$ and $R_Z$ are linear operators acting on the other $\omega\setminus\{j\}$ qubits.
Since $U_{\omega}\rho_{\omega}U_{\omega}^{\dagger}=\rho_{\omega}$, we have $U_jZ_jU_j^{\dagger}=\pm Z_j$.
\end{IEEEproof}

The following corollary about the elements of the automorphism group of a stabilizer code is immediate from Lemma~\ref{lem:ujnsupp}. After this work was completed, we learned that the same statement was independently obtained by D.~Gross and M.~Van den Nest \cite{gross07} and that the theorem was first proved in the diploma thesis of D.~Gross \cite{grossthesis}.

\begin{corollary} \label{cor:form} $U\in\aut{(Q)}$ for a stabilizer code $Q$ iff
\begin{equation}\label{eq:form}
U=L\left( \bigotimes_{j=1}^n \diag{(1,e^{i\theta_j})}\right) R^\dag
P_\pi\in {\mathbb V}(Q)
\end{equation}
for some local Clifford unitaries $L=L_1\otimes\dots\otimes L_n$,
$R=R_1\otimes\dots\otimes R_n$, product of swap unitaries $P_\pi$
enacting the coordinate permutation $\pi$, and angles
$\{\theta_1,\dots,\theta_n\}$.
\end{corollary}

\section{Transversality versus Universality}\label{sec:transversal}

In this section we prove that there is no universal set of
transversal gates for binary stabilizer codes.

\begin{definition}\label{def:encuniv}
A set $A\subseteq {\mathbb V}(Q^{\otimes n})$ is \textit{encoded
computationally universal} if, for any $n$, given $U\in {\mathbb
V}(Q^{\otimes n})$,
\begin{equation}
\forall \epsilon>0, \exists V_1,\dots,V_{\eta(\epsilon)}\in A,\
\textrm{s.t.}\ ||UP_Q^{\otimes n}-\left(\prod_i V_i\right)P_Q^{\otimes
n}||<\epsilon.
\end{equation}
Gates in $A$ may be implicitly mapped to isometries on the
appropriate Hilbert space, as in Definition~\ref{def:universality}.
\end{definition}

\begin{theorem}\label{thm:maintheorem}
For any stabilizer code $Q$ that is free of Bell pairs and trivially encoded qubits,
and for all $r\geq 1$, $\trans(Q^{\otimes r})$ is not an encoded computationally universal
set of gates for even one of the logical qubits of $Q$.
\end{theorem}

\begin{IEEEproof}
We prove this theorem by contradiction. We first assume that we can
perform universal quantum computation on at least one of the qubits encoded into $Q$
using only transversal gates.
Then, we pick an arbitrary minimum weight element
$\alpha\in\mathcal{C}(\mathcal{S})\setminus\mathcal{S}$, and perform
appropriate transversal logical Clifford operations on $\alpha$.
Finally, we will identify an element in
$\mathcal{C}(\mathcal{S})\setminus\mathcal{S}$ that has support
strictly contained in $\supp(\alpha)$. This contradicts the fact
that $\alpha$ is a minimal weight element in
$\mathcal{C}(\mathcal{S})\setminus\mathcal{S}$, i.e. that the code
has the given distance $d$.

We first prove the theorem for A) the single block case and then
generalize it to B) the multiblock case.

\subsection{The single block case ($r=1$)}

The first problem we encounter is that general transversal gates,
even those that implement logical Clifford gates, might not map
logical Pauli operators back into the Pauli group. This behavior
potentially takes us outside the stabilizer formalism.

\begin{definition}
The \textit{generalized stabilizer} ${\mathcal I}(Q)$ of a quantum
code $Q$ is the group of all unitary operators that fix the code
space, i.e.
\begin{equation*}
{\mathcal I}(Q)=\{ U\in SU(2^n)\ |\ U|\psi\rangle = |\psi\rangle,\
\forall |\psi\rangle\in Q\}.
\end{equation*}\label{GS}
\end{definition}

The transversal $T$ gate on the 15-qubit Reed-Muller code is one
example of this problem since it maps $\bar{X}=X^{\otimes 15}$ to an
element $(\frac{1}{\sqrt{2}}(X-Y))^{\otimes 15}$. This element is a
representative of $\frac{1}{\sqrt{2}}(\bar{X}+\bar{Y})$ but has many
more terms in its expansion in the Pauli basis. These terms result
from an operator in the generalized stabilizer $\mathcal{I}$.

The 9-qubit Shor code gives another example. A basis for this code
is
\begin{equation}
|0/1\rangle \propto (|000\rangle+|111\rangle)^{\otimes 3}\pm
(|000\rangle-|111\rangle)^{\otimes 3},
\end{equation}
from which it is clear that $e^{i\theta Z_1}e^{-i\theta Z_2}\in
\mathcal{I}(Q_{\textrm{Shor}})\setminus S$. This gate does not map
$\bar{X}=X^{\otimes 9}$ back to the Clifford group, even though it
is both transversal and logically an identity gate in the logical
Clifford group.

In spite of these possibilities, we will now see that we can avoid
further complication and stay within the powerful stabilizer
formalism.

First, we review a well-known fact about stabilizer codes.

\begin{lemma}\label{lem:proj}
Let $\mathcal{S}=\langle M_1,\dots,M_{n-k}\rangle$ be the stabilizer
of an $[[n,k,d]]$ code $Q$. For any $n$-qubit Pauli operator
$R\notin \mathcal{C}(\mathcal{S})$, we have $P_QRP_Q=0$ where $P_Q$
is the projector onto the code subspace.
\end{lemma}

\begin{IEEEproof}
We have
\begin{align*}
P_Q & \propto\prod\limits_{i=1}^{n-k}(I+M_i)\ \textrm{and} \\
P_QR & \propto R\prod\limits_{i=1}^{n-k}(I+(-1)^{r(i)}M_i),
\end{align*}
where $r(i)=0$ if $R$ commutes with $M_i$ and $r(i)=1$ if $R$
anticommutes with $M_i$. $R\notin\mathcal{C}(\mathcal{S})$ so
$r(i)=1$ for at least one $i$, and $(I-M_i)(I+M_i)=0$, which gives
$P_QRP_Q=0$.
\end{IEEEproof}

\begin{lemma}\label{lem:li}
Let $Q$ be a stabilizer code with stabilizer $\mathcal{S}$ and let
$\alpha\in \mathcal{C}(\mathcal{S})\setminus\mathcal{S}$ be a
minimum weight element in
$\mathcal{C}(\mathcal{S})\setminus\mathcal{S}$. Without loss of
generality, $\alpha\in\bar{X}_1\mathcal{S}$ (the $\bar{X}_1$ coset of $\mathcal{S}$), where the subscript indicates 
what logical qubit the logical operator acts on.
If the logical Clifford operations $\bar{H}_1$ and $\bar{P}_1$ on the first encoded qubit are
transversal, then there exists
$\beta,\gamma\in\mathcal{C}(\mathcal{S})\setminus\mathcal{S}$ such
that $\beta\in\bar{Z}_1\mathcal{S}$, $\gamma\in\bar{Y}_1\mathcal{S}$
and $\supp(\alpha)=\supp(\beta)=\supp(\gamma)$.
\end{lemma}

\begin{IEEEproof}
$\bar{H}_1$ is transversal, so
$\beta'':=\bar{H}_1\alpha\bar{H}_1^\dag\in\bar{Z}_1{\mathcal I}$ and
$\xi:=\supp(\alpha)=\supp(\beta'')$. Expand $\beta''$ in the basis
of Pauli operators
\begin{equation}\label{eq:expansion}
\beta'' = \sum_{R\in C(\mathcal{S}), \supp(R)\subseteq\xi} b_R R +
\sum_{R'\in G_n\setminus C(\mathcal{S})} b_{R'} R'
\end{equation}
where $b_R,b_{R'}\in {\mathbb C}$. By Lemma~\ref{lem:proj}, $b_R\neq
0$ for at least one $R\in C(\mathcal{S})$ in the first term of
Eq.~\ref{eq:expansion}. The operator
$\beta':=P_Q\beta''P_Q\in\bar{Z}_1{\mathcal I}$ is a linear
combination of elements of $C(\mathcal{S})$,
\begin{equation}
\beta' = \sum_{R\in C(\mathcal{S})\setminus \mathcal{S},
\supp(R)=\xi} b_R R + \sum_{R'\in \mathcal{S},
\supp(R')\subseteq\xi} b_{R'} R',
\end{equation}
where the terms $R\in C(\mathcal{S})\setminus \mathcal{S}$ must have
support $\xi$ since $\alpha$ has minimum weight in
$C(\mathcal{S})\setminus \mathcal{S}$. Considering the action of
$\beta'$ on a basis of $Q$, it is clear that there is a term
$b_\beta \beta$ where $b_\beta\neq 0$,
$\beta\in\bar{Z}_1\mathcal{S}$, and $\supp(\beta)=\xi$.

Similarly, since $\bar{P}_1$ is transversal, there must exist
$\gamma\in\bar{Z}_1\mathcal{S}$, and $\supp(\gamma)=\xi$.
\end{IEEEproof}

\begin{remark}
Note in the proof of the above lemma, we assume that $\bar{H}_1$ is
exactly transversal, i.e. $\epsilon=0$ in
Definition~\ref{def:encuniv}. However, the proof is also valid for
an arbitrarily small $\epsilon>0$. Indeed, in this case
$\beta''\notin\bar{Z}_1{\mathcal I}$, but $\beta''$ must have a
non-negligible component in $\bar{Z}_1{\mathcal I}$ to approximate
$\bar{H}_1$. Hence, when
expanding $\beta''$ in the Pauli basis, there must exist a
$\beta\in\bar{Z}_1{\mathcal S}$ such that $\supp(\beta)=\xi$, i.e.
the same argument holds even for an arbitrarily small $\epsilon>0$.
\end{remark}

\begin{remark}
The choice of $\alpha\in\bar{X}_1\mathcal{S}$ is made without loss
of generality, since for a given stabilizer code, we have the
freedom to define logical Pauli operators, and this freedom can be
viewed as a ``choice of basis''. What is more, since we assume
universal quantum computation can be performed transversally on the
code, then no matter what basis (of the logical Pauli operators) we
choose, $\bar{H}_1$ and $\bar{P}_1$ must be transversal. On the other hand,
sometimes we would like to fix our choice of basis, as in the case
of a subsystem code, to clearly distinguish some logical qubits
(protected qubits) from other logical qubits (gauge qubits). In this
case, we can choose $\alpha$ as a minimum weight element in
$\{\bar{X}_s\mathcal{S},\bar{Y}_s\mathcal{S},\bar{Z}_s\mathcal{S}\}$,
where $s$ is a distinguished logical qubit. Starting from this
choice of $\alpha$, one can see that the arguments hold for
subsystem codes as well as subspace codes, because the distance of the
subsystem code is defined with respect to this subgroup.
\end{remark}

\begin{remark}
The procedure of identifying $\beta\in\bar{Z}_1{\mathcal S}$ from
$\beta''\in\bar{Z}_1{\mathcal I}$ in the proof of Lemma~\ref{lem:li} is general in the following sense.
We can begin with a minimum weight element of $\alpha\in
\bar{X}_1\mathcal{S}\subset C(\mathcal{S})\setminus \mathcal{S}$ and
apply any transversal logical Clifford gate to generate a
representative $\beta\in C(\mathcal{S})\setminus \mathcal{S}$ of the
corresponding logical Pauli operator such that
$\supp(\alpha)=\supp(\beta)$. This procedure is used a few times in
our proof, so we name this procedure the
``$\mathcal{I}\rightarrow\mathcal{S}$ procedure''.
\end{remark}

Now we can begin the proof of Theorem~\ref{thm:maintheorem}. Assume
that $\trans{(Q)}$ is encoded computationally universal. Let
$\alpha\in \bar{X}_1\mathcal{S}\subset
C(\mathcal{S})\setminus\mathcal{S}$ be a minimum weight element in
$C(\mathcal{S})\setminus\mathcal{S}$. Applying the
``$\mathcal{I}\rightarrow\mathcal{S}$ procedure'' to both
$\bar{H}_1$ and $\bar{P}_1$, we obtain
$\beta\in\bar{Z}_1\mathcal{S}$ and $\gamma\in\bar{Y}_1\mathcal{S}$
such that $\supp(\alpha)=\supp(\beta)=\supp(\gamma)=:\xi$ and
$|\xi|=d$. The next lemma puts these logical operators into a simple
form for convenience.

\begin{lemma}\label{lem:XYZ}
If $\alpha\in\bar{X}_1\mathcal{S}$, $\beta\in\bar{Z}_1\mathcal{S}$,
and $\gamma\in\bar{Y}_1\mathcal{S}$, all have the same support
$\xi$, and $|\xi|=d$ is the minimum distance of the code, then there
exists a local Clifford operation that transforms $\alpha$,
$\gamma$, and $\beta$ to $X^{\otimes |\xi|}$,
$(-1)^{|\xi|/2}Y^{\otimes |\xi|}$, and $Z^{\otimes |\xi|}$,
respectively.
\end{lemma}

\begin{IEEEproof}
Let $\xi=\{i_1,i_2,\ldots,i_{|\xi|}\}$ and write
$\alpha=\alpha_{i_1}\alpha_{i_2}\ldots\alpha_{i_{|\xi|}}$,
$\beta=\beta_{i_1}\beta_{i_2}\ldots\beta_{i_{|\xi|}}$, where each
$\alpha_{i_k}$ and $\beta_{i_k}$, $k\in[|\xi|]$, are one of the
three Pauli matrices $X_{i_k}$,$Y_{i_k}$, or $Z_{i_k}$, neglecting
phase factors $\pm i$ or $-1$.

Apart from a phase factor, $\alpha_{i_k}\neq \beta_{i_k}$ for all
$k\in [|\xi|]$. Indeed, if for some $k$, $\alpha_{i_k}=\beta_{i_k}$,
then $\supp(\bar{Y})\neq\xi$, a contradiction.

Therefore, for each $k\in [|\xi|]$ there exists a single qubit
Clifford operation $L_{i_k}\in {\mathcal L}_1$ such that
$L_{i_k}\alpha_{i_k}L_{i_k}^\dag=X_{i_k}$ and
$L_{i_k}\beta_{i_k}L_{i_k}^\dag=Z_{i_k}$. The local Clifford
operation
\begin{equation}
L_{\xi}=\bigotimes_{k=1}^{j}L_{i_k}
\end{equation}
applies the desired transformation.
\end{IEEEproof}

Applying Lemma~\ref{lem:XYZ}, we obtain a local Clifford operation
that we apply to $Q$. Now we have a locally Clifford equivalent code
$Q'$ for which $\supp{\alpha}=\supp{\beta}=\supp{\gamma}=\xi$ and
$\alpha=X^{\otimes |\xi|}\in\bar{X}_1\mathcal{S}$,
$\gamma=(-1)^{|\xi|/2}Y^{\otimes |\xi|}\in\bar{Y}_1\mathcal{S}$ and
$\beta=Z^{\otimes |\xi|}\in\bar{Z}_1\mathcal{S}$.

Note that if $|\xi|=d$ is even, then the validity of
Lemma~\ref{lem:XYZ} already leads to a contradiction since
$\alpha, \beta, \gamma$ must anti-commute with each other. However,
if $|\xi|=d$ is odd, we need to continue the proof.

By Theorem~\ref{thm:d2lulc} and Lemma~\ref{lem:ujnsupp}, if
$U=\bigotimes_{j=1}^n U_j$ is a transversal gate on one block,
then either $U_j\in {\mathcal L}_1$ for all $j\in\xi$ or
$U_j=Le^{i\theta R}$ for one or more $j\in\xi$, where 
$L\in {\mathcal L}_1$, $\theta\in {\mathbb R}$, and $R\in {\cal G}_1$.
If $U_j\in {\mathcal L}_1$ for all $j\in\xi$ then, for the
first encoded qubit of $Q'$, the only transversal operations are
logical Clifford operations.

Therefore, there must exist a coordinate $j\in\xi$, such that 
$U_j=e^{i\theta Z}$ up to a local Clifford operation.
Since $\bar{H}_1$ is transversal, when expanding $\bar{H}_1\beta\bar{H}_1^{\dagger}\in\bar{X}_1\mathcal{I}$
in the basis of Pauli operators, using the
``$\mathcal{I}\rightarrow\mathcal{S}$ procedure'', we know that
there exists $\alpha'\in\bar{X}_1\mathcal{S}$ and
$\supp(\alpha')=\xi$. Furthermore, since
$(\bar{H}_1)_jZ(\bar{H}_1)_j^{\dagger}=\pm Z$, we have
$(\alpha')_j=Z_j$, i.e. $\alpha'$ restricted to the $j$th qubit is
$Z_j$. We know $\gamma'=i\alpha'\beta\in\bar{Y}_1\mathcal{S}$, and
$(\gamma')_j=I_j$. Therefore, $\supp(\gamma')$ is strictly contained
in $\xi$. However, this contradicts the fact that $\alpha$ is a
minimal weight element in
$\mathcal{C}({\mathcal{S}})\setminus\mathcal{S}$. This concludes the
proof of Theorem~\ref{thm:maintheorem} for the single block case.

\subsection{The multiblock case ($r>1$)}\label{subsec:multiblock}

Now we consider the case with $r$ blocks. A superscript
$(i)$, $i\in [r]$, denotes a particular block. For example,
$U^{(i)}$ acts on the $i$th block. First, we
generalize Theorem~\ref{thm:d2lulc} and Lemma~\ref{lem:ujnsupp} to the
multiblock case.

\begin{lemma}\label{lem:bigblocklem}
Let $Q$ be an $[[n,k,d]]$ stabilizer code free of Bell pairs and trivially encoded qubits,
and let $U$ be a transversal gate on $Q^{\otimes r}$. Then for each $j\in [n]$ either 
$U_j\in {\cal L}_r$ or $U_j=L_1VL_2$ where $L_1$,$L_2\in {\cal L}_1^{\otimes r}$ are
local Clifford gates and $V$ either normalizes the group $\langle\pm Z_j^{(i)}, i\in [r]\rangle$,
of Pauli $Z$ operators or keeps the linear span of its group elements invariant.
\end{lemma}

\begin{IEEEproof}
Lemma~\ref{lem:rains} and Lemma~\ref{lem:Aw} can be generalized to the the multiblock 
case with almost the same proof, which we do not repeat here. In the multiblock case, 
the corresponding results of Lemma~\ref{lem:Aw} read
\begin{eqnarray*}
A_{\omega}(Q)&=&1\ :\
S_{\omega}^{\otimes{r}}(Q)=\{I^{\otimes\omega},
\ Z^{\otimes\omega}\}^{\otimes{r}}\nonumber\\
A_{\omega}(Q)&=&3\ :\nonumber\\
S_{\omega}^{\otimes{r}}(Q)&=&\{I^{\otimes\omega},\
X^{\otimes\omega}, (-1)^{(|\omega|/2)}Y^{\otimes\omega},
Z^{\otimes\omega}\}^{\otimes r}\label{MEr},
\end{eqnarray*}
and the corresponding equation of Eq.~\ref{Uomega} in Lemma~\ref{lem:rains} is
\begin{equation}
U_{\omega}\rho_{\omega}^{\otimes r}(Q')U_{\omega}^{\dagger}=\rho_{\omega}^{\otimes r}(Q).
\end{equation}

When $A_{\omega}(Q)=3$, we need to generalize the result of Lemma~\ref{lem:d2LC}.
In particular, if $U=\bigotimes_{i=1}^r U_j\in U(2^r)^{\otimes 2m}$
satisfies $U\rho_{[[2m,2m-2,2]]}^{\otimes r}U^\dag=\rho_{[[2m,2m-2,2]]}^{\otimes r}$,
then for each $j\in\omega$, $U_j\in U(2^r)$ is a Clifford operator. 
Indeed, any $r$-qubit unitary operator $V\in U(2^r)$ acts on a Pauli operator
$\sigma_{a_1}\sigma_{a_2}\dots\sigma_{a_r}$ as
\begin{align*}
\sigma_{a_1}\sigma_{a_2}\dots\sigma_{a_r} & \mapsto V\sigma_{a_1}\sigma_{a_2}\dots\sigma_{a_r}V^\dag \\
& = \sum_{i_1,...i_r=0}^{3} o_{a_1\dots a_ri_1\dots i_r}\sigma_{i_1}\sigma_{i_2}\dots\sigma_{i_r},
\end{align*}
for every nonidentity Pauli string $a$, where $(o_{a,i_1,...i_r})\in SO(4^r-1)$ and 
$o_{a,0,\dots,0}=0$.
We can rearrange the numbering of the coordinates in
$\rho_{[[2m,2m-2,2]]}^{\otimes r}$ such that the coordinate
$r(a-1)+b$ denotes the $a$th qubit of the $b$th block. In the
standard basis $\{|0\rangle,...,|4^r-2\rangle\}$ of
${\mathbb R}^{4^r-1}$, $\rho_{[[2m,2m-2,2]]}^{\otimes r}$
is associated to the vector
\begin{equation}
v := \sum_{j=0}^{4^r-2}{b_j}|jj\dots j\rangle\in ({\mathbb R}^{4^r-1})^{\otimes 2m}
\end{equation}
where $b_j\in\{\pm 1\}$. The vector is acted on by orthogonal matrices in $SO(4^r-1)^{\otimes 2m}$.
For a code free of Bell pairs, we have $|\omega|\geq 4$.
By reasoning similar to the proof of Lemma~\ref{lem:d2LC}, if
$O=O_1\otimes\dots\otimes O_{2m}\in SO(4^r-1)^{\otimes 2m}$
satisfies $Ov=v$, then each $O_i$ is a monomial matrix. This
implies that $U_j\in U(2^r)$ is a Clifford operator for each $j\in\omega$.

When $A_{\omega}(Q)=1$, it is possible to follow reasoning similar to the proof of Theorem~\ref{thm:d2lulc}.
Now the equations analogous to Eqs.~\ref{eq:Aw11} are
\begin{align}
\rho_\omega(Q)^{\otimes r} & = \left(\frac{1}{2^{|\omega|}}(I^{\otimes|\omega|}+M_\omega)\right)^{\otimes r}\label{eq:rhoromega} \\
\rho_\mu(Q)^{\otimes r} & =
\left(\frac{1}{2^{|\mu|}}(I^{\otimes|\mu|}+N_\mu)\right)^{\otimes
r}.\label{eq:rhormu}
\end{align}
Up to local Clifford operations, we can choose $M_\omega=X^{\otimes|\omega|}$ and $N_\mu=Z^{\otimes|mu|}$

We again rearrange the numbering of the coordinates of $\rho_\omega(Q)^{\otimes r}$ such that the 
coordinate $r(a-1)+b$ denotes the $a$th qubit of the $b$th block. In the standard
basis $\{|0\rangle,|1\rangle,...,|2^r-2\rangle\}$ of ${\mathbb R}^{2^r-1}$, the matrix 
$\rho_\omega(Q)^{\otimes r}$ is associated to the vector
\begin{equation}
v := \sum_{j=0}^{2^r-2}|jj\dots j\rangle\in ({\mathbb R}^{2^r-1})^{\otimes |\omega|}\label{eq:vr}
\end{equation}
acted on by $SO(2^r-1)^{\otimes |\omega|}$. 

Note for any coordinate $j\in[n]$, if there are elements $R_1,R_2,R_3\in\mathcal{M}(Q)$
such that $(R_1)_j=X$, $(R_2)_j=Y$, and $(R_3)_j=Z$, then we have both $|\omega|\geq 2$ and 
$|\mu|\geq 2$ \cite{zeng:lulc}. Following similar reasoning to the proof of Lemma~\ref{lem:d2LC},
if $O=O_1\otimes\dots\otimes O_{|\omega|}\in SO(2^r-1)^{\otimes |\omega|}$ satisfies $Ov=v$, then
each $O_i$ has a monomial subblock. Therefore, $U_j\in\mathcal{N}(\langle \pm X_j^{(i)}\rangle_{i=1}^{r})$
for each $j\in\omega$, where $\mathcal{N}(\langle\pm X_j^{(i)}\rangle_{i=1}^{r})$ is normalizer of the
group $\langle\pm X_j^{(i)}\rangle_{i=1}^{r}$ of Pauli $X$ operators acting at the $j$th coordinate
of the $i$th block.
Similarly for $\rho_\mu(Q)^{\otimes r}$, $U_j\in\mathcal{N}(\langle\pm Z_j^{(i)}\rangle_{i=1}^{r})$ for
each $j\in\mu$, where $\mathcal{N}(\langle\pm Z_j^{(i)}\rangle_{i=1}^{r})$ is normalizer of the group
$\langle\pm Z_j^{(i)}\rangle_{i=1}^{r}$ of Pauli $Z$ operators.
$\mathcal{N}(\langle\pm X_j^{(i)}\rangle_{i=1}^{r})\cap\mathcal{N}(\langle\pm Z_j^{(i)}\rangle_{i=1}^{r})$ 
is a subgroup of the Clifford group, therefore $U_j\in U(2^r)$ is a Clifford operator
for all $j\in\omega\cap\mu$.
If instead $(R_a)_j=(R_b)_j$ for all $R_a,R_b\in\mathcal{M}(Q)$ and if $|\omega|\geq 3$, then
$U_j\in\mathcal{N}(\langle \pm Z_j^{(i)}\rangle_{i=1}^{r})$ up to local Clifford operations, but $U_j$ is
not necessarily a Clifford operator.

If $|\omega|=2$ and $A_{\omega}=1$, then the form
of the vector in Eq.~\ref{eq:vr} leads to different behavior when $r>1$. When $r=1$, there is only
one term in the summation, so $U_j\in\mathcal{N}(\langle\pm X_j\rangle)$. However, when $r>1$,
generally $U_j\notin\mathcal{N}(\langle\pm X_j^{(i)}\rangle_{i=1}^{r})$. Nevertheless $U_j$ must keep
$\text{span}(\langle\pm X_j^{(i)}\rangle_{i=1}^{r})$ invariant, where
$\text{span}(\langle\pm X_j^{(i)}\rangle_{i=1}^{r}\rangle)$ is the space of linear operators
spanned by the group $\langle\pm X_j^{(i)}\rangle_{i=1}^{r}$ with coefficients in $\mathbb C$. 
Indeed, consider Eq.~\ref{eq:vr} when $|\omega|=2$. For convenience, let $\omega=\{1,2\}$. We have 
$\text{Tr}_2(vv^T)=\sum_{j=1}^r|j\rangle\langle j|$. If $Ov=v$, then 
$\text{Tr}_2(Ovv^TO^T)=\sum_{j=1}^r|j\rangle\langle j|$. However, 
\begin{align*}
\text{Tr}_2(Ovv^TO^T)& =\text{Tr}_2(O_1vv^TO_1^T) \\
& =\sum_{j=1}^rO_1|j\rangle\langle j|O_1^T \\
& =\sum_{k,k'}(\sum\limits_{j=1}^r(O_1)_{jk}(O_1)_{jk'})|k\rangle\langle k'|. 
\end{align*}
Therefore $\sum_{j=1}^r(O_1)_{jk}(O_1)_{jk'}=\delta_{kk'}$ for all $k,k'\leq r$ 
and $\sum_{j=1}^r(O_1)_{jk}(O_1)_{jk'}=0$ for any one of $k>r$ or $k'>r$, which 
means $(O_1)_{jk}=0$ for all $k>r$.

Finally, we need to generalize Lemma~\ref{lem:ujnsupp} to the multiblock case. Recalling Eq.~\ref{eq:rhonsupp}, 
we now have 
\begin{equation}
\rho_\omega=\left(\frac{1}{2^{|\omega|}}(I_j\otimes R_I+Z_j\otimes R_Z)\right)^{\otimes r}. 
\end{equation}
This projector can be associated with a vector $v=\sum_{j=1}^r|jj\rangle$. Using the same technique 
as for the case where $A_\omega=1$ and $|\omega|=2$, we conclude that $U_j$ must keep one of the three
spaces of linear operators $\text{span}(\langle\pm X_j^{(i)}\rangle_{i=1}^{r})$, 
$\text{span}(\langle\pm Y_j^{(i)}\rangle_{i=1}^{r})$, or $\text{span}(\langle\pm Z_j^{(i)}\rangle_{i=1}^{r})$ invariant.
\end{IEEEproof}

Given these generalizations of Theorem~\ref{thm:d2lulc} and Lemma~\ref{lem:ujnsupp} to the multiblock case, we
now show that all the arguments in the proof of the single block
case can be naturally carried to the multiblock case. Most
importantly, we show that the
``$\mathcal{I}\rightarrow\mathcal{S}$'' procedure is still valid. To
specify the ``$\mathcal{I}\rightarrow\mathcal{S}$'' procedure for
the multiblock case, we first need to generalize the concept of the 
generalized stabilizer defined in Definition~\ref{GS} to the multiblock case.

\begin{definition}
The \textit{generalized stabilizer} ${\mathcal I}(Q^{\otimes r})$ of
an $r$-block quantum code $Q^{\otimes r}$ is the group of all unitary
operators that fix the code space, i.e.
\begin{equation*}
{\mathcal I}(Q^{\otimes r})=\{ U\in SU(2^{nr})\ |\ U|\psi\rangle =
|\psi\rangle,\ \forall |\psi\rangle\in Q^{\otimes r}\}.
\end{equation*}
\end{definition}

Similar to the single block case, we start by assuming that
universal quantum computation can be performed using transversal gates.
Then $\bar{H}_1^{(1)}$, the logical Hadamard operator acting on the 
first logical qubit of the first block, is transversal.

Let $\alpha^{(1)}$ be a minimal weight element of
$\mathcal{C}(\mathcal{S})\setminus\mathcal{S}$. Without loss of
generality, we assume $\alpha^{(1)}\in\bar{X}_1\mathcal{S}$. Then
$\bar{H}_1^{(1)}$ will transform $\alpha^{(1)}$ to some
$\beta'\in\bar{Z}_1^{(1)}\mathcal{I}(Q^{\otimes r})$.
This is to say, $\beta'$ acting on $P_Q^{\otimes r}$ is a logical
$Z$ operation on the first logical qubit of the first block, and
identity on the other $r-1$ blocks. However, this does not
mean that $\beta'=\beta^{''(1)}\bigotimes\limits_{i=2}^{r}\delta^{'(i)}$,
where $\beta''\in\bar{Z}_1\mathcal{S}$ and
$\delta'^{(i)}\in\mathcal{I}(Q)$ for all $i=2,\dots,r$, because
\begin{equation}
\bar{H}_1^{(1)}=\bigotimes_{j=1}^{n} U_j,
\end{equation}
where each $U_j$ acts on $r$ qubits.

Expanding $\beta'$ in the basis of $nr$ qubit Pauli operators.
For the same reason shown in the proof of Lemma~\ref{lem:li}, there 
must be at least one term in the expansion which has the form
\begin{equation}
\beta^{(1)}\bigotimes\limits_{i=2}^{r}\delta^{(i)},
\end{equation}
where $\beta\in\bar{Z}_1\mathcal{S}$, and $\delta^{(i)}\in\mathcal{S}$ for
all $i=2,\dots,r$.
So, the generalization of the ``$\mathcal{I}\rightarrow\mathcal{S}$
procedure'' to the multiblock case is clear: Pauli operators acting on
a code are either logical Pauli operators (on any number of
qubits and any number of blocks) or they map the code $P_Q^{\otimes
r}$ to an orthogonal subspace. Nevertheless, this is an important observation.

Now $\beta\in\mathcal{C}(\mathcal{S})\setminus\mathcal{S}$ is a
logical $Z$ operation acting on the first logical qubit of a single
block of the code. Due to Eq.~\ref{eq:trans}, $\supp(\beta^{(1)})\subseteq\supp(\alpha^{(1)})$. 
However $\alpha^{(1)}$ is a minimal weight element of
$\mathcal{C}(\mathcal{S})\setminus\mathcal{S}$, therefore we have
$\supp(\beta^{(1)})=\supp(\alpha^{(1)}):=\xi$. For convenience, we now
drop the superscript $(1)$ when referring to these logical operators.

Since $\bar{P}_1^{(1)}$ is transversal, then there exists
a $\gamma\in\bar{Y}_1\mathcal{S}$ which has the same support as
$\alpha$.
Now we have $\alpha\in\bar{X}_1\mathcal{S}$,
$\beta\in\bar{Z}_1\mathcal{S}$ and $\gamma\in\bar{Y}_1\mathcal{S}$
such that $\supp(\gamma)=\supp(\beta)=\supp(\alpha)$.

Like the single block case, by Lemma~\ref{lem:XYZ}, there is a local Clifford
operation such that $\alpha=X^{\otimes |\xi|}\in\bar{X}_1\mathcal{S}$,
$\beta=(-1)^{|\xi|/2}Y^{\otimes |\xi|}\in\bar{Y}_1\mathcal{S}$ and
$\gamma=Z^{\otimes |\xi|}\in\bar{Z}_1\mathcal{S}$. If $|\xi|=d$ is even
there is a contradiction, since $\alpha$, $\beta$, $\gamma$ must anti-commute 
with each other. 

When $|\xi|=d$ is odd, we need the following arguments.
If for all coordinates $j\in\xi$, there are elements $R_1,R_2,R_3\in\mathcal{M}(Q)$ 
such that $(R_1)_j=X$, $(R_2)_j=Y$ and $(R_3)_j=Z$, then
$U_j\in U(2^r)$ is a Clifford operator for all $j\in\xi$ by Lemma~\ref{lem:bigblocklem}. 
Therefore, all the possible logical operations that are transversal on the first 
encoded qubit must be Clifford operations, contradicting the assumption that
universality can be achieved by transversal gates.

Therefore,  there exists a coordinate $j\in\xi$ such that either (a) there is no
minimal support containing $j$ or (b) $(R_1)_j=(R_2)_j\neq I$ for all $R_1,R_2\in \mathcal{M}(Q)$.
Use the ``$\mathcal{I}\rightarrow\mathcal{S}$ procedure'' to
expand $\bar{H}_1^{(1)}\beta(\bar{H}_1^{(1)})^{\dagger}$ in the
basis of Pauli operators and extract $\alpha'\in\bar{X}_1\mathcal{S}$ with $\supp(\alpha')=\xi$.
Since $\bar{H}_1$ is transversal, $U_j$ must keep 
$\text{span}(\langle\pm Z_j^{(i)}\rangle_{i=1}^{r})$ invariant, up to a local Clifford operation. 
This means that the $j$th coordinate of $\alpha'$ is either $I_j$ or $Z_j$. The former is
not possible since $\supp(\alpha')=\xi$. For the later,
$\gamma''=i\alpha'\beta\in\bar{Y}_1\mathcal{S}$, and the $j$th coordinate of $\gamma''$
is $I_j$. Therefore, $\supp(\gamma'')$ is strictly contained in $\xi$. However, this
contradicts the fact that $\alpha$ is a minimal weight element in
$\mathcal{C}({\mathcal{S}})\setminus\mathcal{S}$.
\end{IEEEproof}

\section{The effect of coordinate permutations}\label{sec:permutation}

In this section we discuss the effect of coordinate permutations.

\begin{theorem}\label{thm:permutation}
For any stabilizer code $Q$ free of trivially encoded qubits, $\aut(Q)$ is not an encoded
computationally universal set of gates for any logical qubit.
\end{theorem}

\begin{IEEEproof}
Choose a minimum weight element
$\alpha\in\mathcal{C}(\mathcal{S})\setminus\mathcal{S}$. Without
loss of generality, assume $\alpha\in\bar{X}_1\mathcal{S}$ and
$\supp(\alpha)=\xi$.

Define a single qubit non-Clifford gate $F$ by
\begin{equation}
F:\ X\rightarrow X'=\frac{1}{\sqrt{3}}\left(X+Y+Z\right);\
Z\rightarrow Z'
\end{equation}
where $Z'$ is any operator that is unitary, Hermitian and
anticommuting with $X'$.
We cannot use the idea of applying $\bar{H}_1$ and $\bar{P}_1$
from within $\aut(Q)$ since they might involve different permutations.
We instead assume the logical gate $\bar{F}_1$ can be approximated to
an arbitrary accuracy by gates in $\aut{(Q)}$.
Then we have
\begin{equation}
\bar{F}_1\alpha\bar{F}_1^{\dagger}=\eta\in\frac{1}{\sqrt{3}}\left(\bar{X}_1+\bar{Y}_1+\bar{Z}_1\right)\mathcal{I}(Q)
\end{equation}

Applying the $\mathcal{I}\rightarrow\mathcal{S}$ procedure to $\eta$,
we find $\alpha'\in{\bar{X}_1}\mathcal{S}$,
$\beta'\in{\bar{Y}_1}\mathcal{S}$, and
$\gamma'\in{\bar{Z}_1}\mathcal{S}$ such that
$\supp{\alpha'}=\supp{\beta'}=\supp{\gamma'}=\xi'$ and
$|\xi'|=|\xi|=d$.
By Lemma~\ref{lem:XYZ}, we can find a locally Clifford equivalent code
such that $\alpha'=X^{\otimes |\xi'|}\in\bar{X}_1\mathcal{S}$, $\beta'=(-1)^{|\xi'|/2}Y^{\otimes
|\xi'|}\in\bar{Y}_1\mathcal{S}$ and $\gamma'=Z^{\otimes
|\xi|}\in\bar{Z}_1\mathcal{S}$. Again, $d$ must be odd.

If for all coordinates $j\in\xi$, there are elements
$R_1,R_2,R_3\in\mathcal{M}(Q)$ such that $(R_1)_j=X$, $(R_2)_j=Y$,
and $(R_3)_j=Z$, then for $U\in\aut(Q)$, $U_j\in U(2)$ is a Clifford operator for all 
$j\in\xi$ by Theorem~\ref{thm:d2lulc}. Permutations are Clifford operations as well, so all possible
transversal logical operations on the first encoded qubit
must be Clifford operations, contradicting the assumption that
the transversal gates are a universal set. 

Therefore, there exists $j'\in\xi'$ such that either (a) only one of $\{X,Y,Z\}$ appears in 
$\mathcal{M}(Q)$ at coordinate $j'$ or (b) there is no minimal element with support at $j'$. 
Without loss of generality, we assume that $X$ appears at coordinate $j'$ in case (a). 
Since $\bar{F}_1$ can be performed via some transversal gate plus permutation, we have
\begin{equation}
\bar{F}_1\alpha'\bar{F}_1^{\dagger}=\eta'\in\frac{1}{\sqrt{3}}\left(\bar{X}_1+\bar{Y}_1+\bar{Z}_1\right)\mathcal{I}(Q).
\end{equation}
Again applying the $\mathcal{I}\rightarrow\mathcal{S}$ procedure to
$\eta'$ we know there exist $\alpha''\in{\bar{X}_1}\mathcal{S}$,
$\beta''\in{\bar{Y}_1}\mathcal{S}$, and
$\gamma''\in{\bar{Z}_1}\mathcal{S}$ such that
$\supp{\alpha''}=\supp{\beta''}=\supp{\gamma''}=\xi''$. And
$|\xi''|=|\xi|=d$. The permutation maps $j'$ to $j''$.
However, we know that $\eta'|_{j''}=X$, and this is also true in case (b) by similar reasoning to
Lemma~\ref{lem:ujnsupp}, hence $\alpha''|_{j''}=\beta''|_{j''}=\gamma''|_{j''}=X$. Then
$\gamma'''=i\alpha''\beta''\in\bar{Z}_1\mathcal{S}$ such that
$i\alpha''\beta''|_{j''}=I$. Therefore, $\supp(\gamma''')$ is strictly contained in $\xi$, which contradicts
the fact that $\alpha$ is a minimal weight element in
$\mathcal{C}(\mathcal{S})\setminus\mathcal{S}$.
\end{IEEEproof}

If $\aut(Q)$ is replaced by $\aut(Q^{\otimes r})$, the theorem still holds
because we can view $Q^{\otimes r}$ as another stabilizer code.
However, it is not a simple generalization to allow permutations between
transversal gates acting on $r>1$ blocks. This is because
permutations are permitted to be different on each block and may also
be performed between blocks.

\section{Applications and Examples}\label{sec:applications}

In this section, we apply the proof techniques we have used in previous sections to reveal more facts about the form of transversal non-Clifford gates. 
First, we describe the form of transversal non-Clifford gates on stabilizer codes. We explore further properties of allowable transversal gates 
in the single block case and discuss how the allowable transversal gates relate to the theory of classical divisible codes. Finally, we review a
{\em CSS} code construction based on Reed-Muller codes that yields quantum codes with various minimum distances and transversal non-Clifford gates.

Corollary~\ref{cor:form} gives a form for an arbitrary stabilizer
code automorphism. Similarly, in the multiblock case, Lemma~\ref{lem:bigblocklem} 
provides possible forms of $U_j$ for any transversal
gate $U=\bigotimes_{j=1}^n U_j$. These forms prevent
certain kinds logical gates from being transversal on a stabilizer
code. 

\begin{corollary}
An $r$-qubit logical gate $U$ such that $U_j\notin {\mathcal L}_r$ for all $j$ is transversal 
on a stabilizer code only if $U$ keeps the operator space $\text{span}(\langle\pm\bar{Z}_i\rangle_{i=1}^{r})$
invariant up to a local Clifford operation\label{cor:formr}. Here, $\bar{Z}_i$ denotes the logical Pauli $Z$ operator on the $i$th encoded qubit.
\end{corollary}

\begin{remark}
This is a direct corollary from Theorem~\ref{thm:maintheorem} in Sec.~\ref{sec:transversal}
and Theorem~\ref{thm:permutation} in Sec.~\ref{sec:permutation}. 
\end{remark}

\begin{example}
Consider the three-qubit bit-flip code with stabilizer
$\mathcal{S}=\{Z_1Z_2,Z_2Z_3\}$, and choose
$|0\rangle_L=|000\rangle,|1\rangle_L=|111\rangle$. The
Toffoli gate is transversal on this code and is given by
$\text{Toffoli}=\bigotimes_{j=1}^3\text{Toffoli}_j$. The
Toffoli gate up to a local Clifford is not in $\mathcal{N}(\langle\pm Z_i\rangle_{i=1}^{3})$; however,
the Toffoli gate up to a local Clifford does keep $\text{span}(\langle\pm Z_i\rangle_{i=1}^{3})$
invariant.
\end{example}

\begin{remark}
If $U_j$ up to a local Clifford keeps $\text{span}(\langle\pm Z_j^{(i)}\rangle_{i=1}^{r})$ invariant, 
i.e. $U_j$ transforms any diagonal matrix to a diagonal matrix, then $U_j$ is a monomial matrix.
Similarly, if $U$ keeps $\text{span}(\langle\pm X_j^{(i)}\rangle_{i=1}^{r})$ $($or
$\text{span}(\langle\pm Y_j^{(i)}\rangle_{i=1}^{r})$$)$ invariant, then $U_j$ is a
monomial matrix in the $X_j$ (or $Y_j$) representation. This does not necessarily mean
that $U=\bigotimes_{j=1}^n U_j$ is a monomial matrix (in one of the $X,Y,Z$ representations) 
in the $2^{nr}$ dimensional Hilbert space, since in general some of the $U_j$ might be Clifford operations. 
\end{remark}

\begin{remark}
Corollary~\ref{cor:formr} also applies to a set of gates. 
A set of gates $V_i$, $i=1,\dots,k$, $(V_i)_j\notin {\mathcal L}_r$ for all $j\in [n]$,
is transversal on a stabilizer code only if
all of the $V_i$ up to the same local Clifford keep the operator space
$\text{span}(\langle\pm Z_i\rangle_{i=1}^{r})$ invariant.
\end{remark}

\begin{example}
The set of gates $\{\text{Hadamard}, \text{Toffoli}\}$ cannot
both be transversal on any stabilizer code, since Hadamard
keeps $\text{span}(\langle\pm Y_i\rangle_{i=1}^{r})$ invariant and Toffoli
keeps $\text{span}(\langle\pm Z_i\rangle_{i=1}^{r})$ invariant. These
observations imply that all transversal gates are Clifford, but Toffoli is not
Clifford. Note
$\{\text{Hadamard}, \text{Toffoli}\}$ is ``universal" for quantum
computation in a sense that all the real gates can be approximated
to an arbitrary accuracy \cite{shi:littlehelp}.
\end{example}

Now we restrict ourselves to the single block case. Up to local
Clifford equivalence, Corollary~\ref{cor:form} and
Corollary~\ref{cor:formr} say that the unitary part of a code
automorphism is a diagonal gate. Therefore, we may restrict our
discussion of the essential non-Clifford elements of $\aut{(Q)}$ to
diagonal gates, because we can imagine considering the diagonal
automorphisms for all locally Clifford equivalent codes and their
permutation equivalent codes to find all of the non-Clifford
automorphisms. We further restrict ourselves to the case where the stabilizer code
is {\em CSS} code.

\begin{lemma} Let $Q$ be a {\em CSS} code $CSS(C_1,C_2)$ constructed from classical binary codes $C_2^\perp <C_1$. Then
\begin{equation}
V = \bigotimes_{\ell=1}^n \diag{(1,e^{i\theta_\ell})}\in\aut{(Q)}
\end{equation}
iff $\forall c,c'\in C_2^\perp$ and $\forall a\in C_1/C_2^\perp$,
\begin{equation}\label{eq:diag}
\sum_{\ell\in\supp{(a+c)}} \theta_\ell = \sum_{\ell\in\supp{(a+c')}}
\theta_\ell\ \textrm{mod}\ 2\pi.
\end{equation}
\end{lemma}

\begin{IEEEproof}
The states
\begin{equation}
|\tilde{a}\rangle \propto \sum_{c\in C_2^\perp} |a+c\rangle, a\in
C_1/C_2^\perp,
\end{equation}
are a basis for $Q$. $V$ is diagonal, so $V|c\rangle=v(c)|c\rangle$
for $c\in C_1$ and a factor $v(c)\in {\mathbb C}$ that is a sum of angles. 
$V$ is a logical operation so $V|\tilde{a}\rangle\in Q$, which is
possible for a diagonal gate iff $v(a+c)=v(a+c')$ for all $a\in C_1/C_2^\perp$ 
and all $c,c'\in C_2^\perp$.
\end{IEEEproof}

We now restrict to the case where the angles $\theta_\ell=\theta$ are all equal.

\begin{corollary} Let $Q$ be a {\em CSS} code constructed from classical binary codes $C_2^\perp <C_1$. A gate $V\in\aut{(Q)}$ is a tensor product
of $n$ diagonal unitaries $V_\theta=\diag{(1,e^{i\theta})}$ iff
$\forall c,c'\in C_2^\perp$ and $\forall a\in C_1/C_2^\perp$,
\begin{equation}
\frac{\theta}{2\pi}(\wt{(a+c)}-\wt{(a+c')})\in {\mathbb Z},
\end{equation}
where $\wt{c}$ denotes the Hamming weight of a classical codeword.
\end{corollary}

The corollary's condition can be satisfied if and only if the weight
of all the codewords in $C_1$ are divisible by a common divisor.

\begin{definition}
A classical linear code is said to be divisible by $\Delta$ if
$\Delta$ divides the weight of each codeword. A classical linear
code is divisible if it has a divisor larger than $1$. An $[n,k]$
classical code can be viewed as a pair $(V,\Lambda)$ where $V$ is a
$k$-dimensional binary vector space and
$\Lambda=\{\lambda_1,\dots,\lambda_n\}$ is a multiset of $n$ members
of the dual space $V^\ast$ that serve to encode $v\in V$ as
$c=(\lambda_1(v),\dots,\lambda_n(v))$ and the image of $V$ in
$\{0,1\}^n$ is $k$-dimensional. The $b$-fold replication of $C$ is
$(V,r\Lambda)$ where $r\Lambda$ is the multiset in which each member
of $\Lambda$ appears $r$ times.
\end{definition}

The following theorem, which is less general than that proven in
\cite{ward:divisible}, gives evidence (though not a proof) that
the allowable value $\theta$ might only be $\frac{\pi}{2^{(k+2)}}$,
which implies $U\in\mathcal{C}_k^{(1)}$ (see
Definition~\ref{def:ck}). It would be interesting if all of the
transversal gates for stabilizer codes lie within the $\mathcal{C}_k$
hierarchy.

\begin{theorem}[\cite{ward:divisible}]
Let $C$ be an $[n,k]$ classical binary code that is divisible by
$\Delta$, and let $b=\Delta/\textrm{gcd}(\Delta,2^{k-1})$. Then $C$
is equivalent to a $b$-fold replicated code, possibly with some
added $0$-coordinates.
\end{theorem}

The Reed-Muller codes are well-known examples of divisible codes.
Furthermore, they are nested in a suitable way and their dual codes
are also Reed-Muller codes, which makes them amenable to the {\em CSS}
construction. In particular:

\begin{theorem}[1.10.1, \cite{huffman:HP}]
Let $RM(r,m)$ be the $r$th order Reed-Muller code with block size
$n=2^m$ and $0\leq r\leq m$. Then
\begin{itemize}
\item[(i)] $RM(i,m)\subseteq RM(j,m)$, $0\leq i\leq j\leq m$
\item[(ii)] $\dim{RM(r,m)} = \sum_{i=0}^r {m\choose i}$
\item[(iii)] $d=2^{m-r}$
\item[(iv)] $RM(m,m)^\perp=\{0\}$ and if $0\leq r<m$ then $RM(r,m)^\perp=RM(m-r-1,m)$.
\end{itemize}
\end{theorem}

\begin{lemma}
$RM(r,m)$ is divisible by $\Delta=2^{\lfloor m/r\rfloor - 1}$.
\end{lemma}

\begin{corollary}
Let $even(RM^\ast(r,m))=C_2^\perp <C_1=RM^\ast(r,m)$ where $0<r\leq
\lfloor m/2\rfloor$. Then $CSS(C_1,C_2)$ is an
$[[n=2^m-1,1,d=\textrm{min}(2^{m-r}-1,2^{r+1}-1)]]$ code with a
transversal gate $G=\otimes_{j=1}^n \diag{(1,e^{i2\pi/\Delta})}$
enacting $\bar{G}=\diag{(1,e^{-i2\pi/\Delta})}\in \mathcal{C}_{\log_2
\Delta}^{(1)}$ where $\Delta=2^{\lfloor m/r\rfloor -1}$.
\end{corollary}

For instance, the $[[2^m-1,1,3]]$ {\em CSS} codes constructed from the
first-order punctured Reed-Muller code $R^\ast(1,m)$ and its even
subcode $even(R^\ast(1,m))$ support the transversal gate
$\exp(-i\frac{\pi}{2^{m-1}}\bar{Z})$ \cite{zeng:lulc,steane:fta}.  The
smallest of these, a $[[15,1,3]]$ mentioned in the introduction, has
found application in magic state distillation schemes
\cite{bravyi:magic} and measurement-based fault-tolerance schemes
\cite{raussendorf:oneway}. If we choose parameters $m=8$ and $r=2$
then we have a $[[255,1,7]]$ code with transversal $T$, but this is
not competitive with the concatenated $[[15,1,3]]$ code. We leave open the 
possibility that other families of classical divisible codes give better {\em CSS} codes 
with $d>3$ or $k>1$ and transversal non-Clifford gates.

\section{Conclusion}

We have proven that a binary stabilizer code with a quantum
computationally universal set of transversal gates for even one of its
encoded logical qubits cannot exist, even when those transversal gates
act between any number of encoded blocks.  Also proven is that even
when coordinate permutations are allowed, universality cannot be
achieved for any single block binary stabilizer code.

To obtain the required contradiction, the proof weaves together
results of Rains and Van den Nest that have been generalized to
multiple encoded blocks. Along the way, we have understood the form of
allowable transversal gates on stabilizer codes, which leads to the
fact that the form of gates in the automorphism group of the code is
essentially limited to diagonal gates conjugated by Clifford
operations, together with coordinate permutations. This observation
suggests a broad family of quantum {\em CSS} codes that can be derived from
classical divisible codes and that exhibit the attainable non-Clifford
single-block transversal gates.  In general, it is not clear how to
systematically find non-Clifford transversal gates, but the results in
Section~\ref{sec:applications} take steps in this direction. It would
be interesting to find more examples of codes with non-Clifford
transversal gates.

There remain some potential loopholes for achieving universal
computation with transversal or almost-transversal gates on binary
stabilizer codes. For example, we could relax the definition of
transversality to allow coordinate permutations on all $nr$ qubits
before and/or after the transversal gate. We could also permit each
block to be encoded in a different stabilizer code, and even allow
gates to take an input encoded in a code $Q_1$ to an output encoded
in a code $Q_2$, provided the minimum distances of these codes are
comparable. We could further relax the definitions of transversality
and conditions for fault-tolerance so that each $U_i$ acts on a
small number of qubits in each block. This latter method is
fault-tolerant provided that each $U_i$ acts on fewer than $t$ qubits.
Finally, the generalization to nonbinary stabilizer codes, and further to
arbitrary quantum codes, remain open possibilities.

\section*{Acknowledgments}

We thank Panos Aliferis, Sergey Bravyi, David
DiVincenzo, Ben Reichardt, Graeme Smith, John Smolin, and Barbara Terhal for
comments, criticisms, and corrections. AC is partially supported by a
research internship at IBM.

\bibliographystyle{IEEEtran}
\bibliography{IEEEabrv,tn_vs_un}

\end{document}